\documentclass[10pt,twocolumn,superscriptaddress,aps]{revtex4}
\usepackage[usenames,dvipsnames]{xcolor}

\usepackage{lscape}
\usepackage{booktabs}
\usepackage{graphicx,color,amssymb}
\usepackage{amsmath}
\usepackage[latin1]{inputenc}
\usepackage[sort&compress]{natbib}
\usepackage[version-1-compatibility]{siunitx}
\newcommand{\proj}[2]{\left| #1 \right> \left< #2 \right|}

\definecolor{darkblue}{rgb}{0,0,.4}
\definecolor{darkred}{rgb}{0.6,0,0}

\newcommand{\abs}[1]{\left|#1\right|}

\newcommand{\ket}[1]{\left| #1 \right>}
\newcommand{\bra}[1]{\left< #1 \right|}
\newcommand{\bracket}[3]{\left< #1 \abs{#2} #3 \right>}

\renewcommand\eqref[1]{Eq.~(\ref{#1})}

\newcommand{\rr}{\mathbf{r}}

\newcommand{\me}[3]{\left< #1 \left| #2 \right| #3 \right>}

\newcommand{\qq}{\mathbf{q}}
\newcommand{\ad}{a^{\dagger}}
\newcommand{\divv}{\nabla\cdot}
\newcommand{\spp}{\sigma^{+}}
\newcommand{\smm}{\sigma^{-}}
\newcommand{\av}[1]{\left<#1\right>}
\newcommand{\I}{\mathrm{i}}

\usepackage[bookmarksopen=true,
citecolor=blue,
colorlinks=true,
urlcolor=blue,
linkcolor=blue]{hyperref}

\begin{document}

\title{Phonon Decoherence of Quantum Dots in Photonic Structures: Broadening of the Zero-Phonon Line and the Role of Dimensionality}

\author{P.~Tighineanu}
\email{petru.tighineanu@mpl.mpg.de}
\affiliation{Niels Bohr Institute,\ University of Copenhagen,\ Blegdamsvej 17,\ 2100 Copenhagen,\ Denmark}
\affiliation{Max Planck Institute for the Science of Light,\ Staudtstra\ss e 2, 91058 Erlangen,\ Germany}
\author{C.~L.~Dree\ss en}
\affiliation{Niels Bohr Institute,\ University of Copenhagen,\ Blegdamsvej 17,\ 2100 Copenhagen,\ Denmark}
\author{C.~Flindt}
\affiliation{Department of Applied Physics,\ Aalto University,\ 00076 Aalto,\ Finland}
\author{P.~Lodahl}
\affiliation{Niels Bohr Institute,\ University of Copenhagen,\ Blegdamsvej 17,\ 2100 Copenhagen,\ Denmark}
\author{A.~S.~S\o rensen}
\email{anders.sorensen@nbi.ku.dk}
\affiliation{Niels Bohr Institute,\ University of Copenhagen,\ Blegdamsvej 17,\ 2100 Copenhagen,\ Denmark}

\begin{abstract}
We develop a general microscopic theory describing the phonon decoherence of quantum dots and indistinguishability of the emitted photons in photonic structures. 
The coherence is found to depend fundamentally on the dimensionality of the structure resulting in vastly different performance for quantum dots embedded in a nano-cavity (0D), waveguide (1D), slab (2D), or bulk medium (3D). 
In bulk, we find a striking temperature dependence of the dephasing rate scaling as $T^{11}$ implying that phonons are effectively 'frozen out' for $T \lesssim \SI{4}{\kelvin}$. 
The phonon density of states is strongly modified in 1D and 2D structures leading to a linear temperature scaling for the dephasing strength. 
The resulting impact on the photon indistinguishability can be important even at sub-Kelvin temperatures. 
Our findings provide a comprehensive understanding of the fundamental limits to photon indistinguishability in photonic structures.
\end{abstract}

\maketitle

Disentangling a quantum system from its fluctuating environment is pivotal to the realization of coherent quantum bits. 
Controlling the sources of noise is particularly challenging in solid-state systems, which contain a myriad of mutually interacting quasi-particles. 
An example is semiconductor quantum dots (QDs), which have proven to be excellent quantum light sources~\cite{lodahl15}.
Two important decoherence mechanisms of QDs are the fluctuating electrostatic~\cite{kuhlmann13,arnold14} and spin~\cite{kuhlmann13,urbaszek13,delteil14} environments but these can be neutralized under appropriate external control~\cite{kuhlmann15,somaschi16,ding16,he13}. 
The electrostatic noise is particularly significant in engineered structures~\cite{majumdar11} but recent experiments have demonstrated how to efficiently suppress it~\cite{kirsanske17,thyrrestrup18}.
The decoherence is then dominated by phonons, the acoustic vibrations of the crystal lattice~\cite{besombes01,muljarov04,grange09,thoma16,zajac16}.
Integrating the QDs into photonic devices is essential for obtaining deterministic and scalable light-matter interfaces~\cite{claudon10,gazzano13,arcari14}.
Such photonic structures also contain a modified \emph{phononic} environment due to the breakdown of translational symmetry.
A unified description of how the modified phonon environment affects the coherence of QDs is lacking despite its vital importance for solid-state quantum optics~\cite{lodahl15}.
Previous founding work concentrated on QDs in bulk media~\cite{muljarov04,grange09}, generic models for 1D and 3D phonon baths~\cite{palma96}, or the special case of linear phonon coupling in nano-wires~\cite{lindwall07} and carbon nanotubes~\cite{galland08}. 

\begin{figure}
\includegraphics[width=\columnwidth]{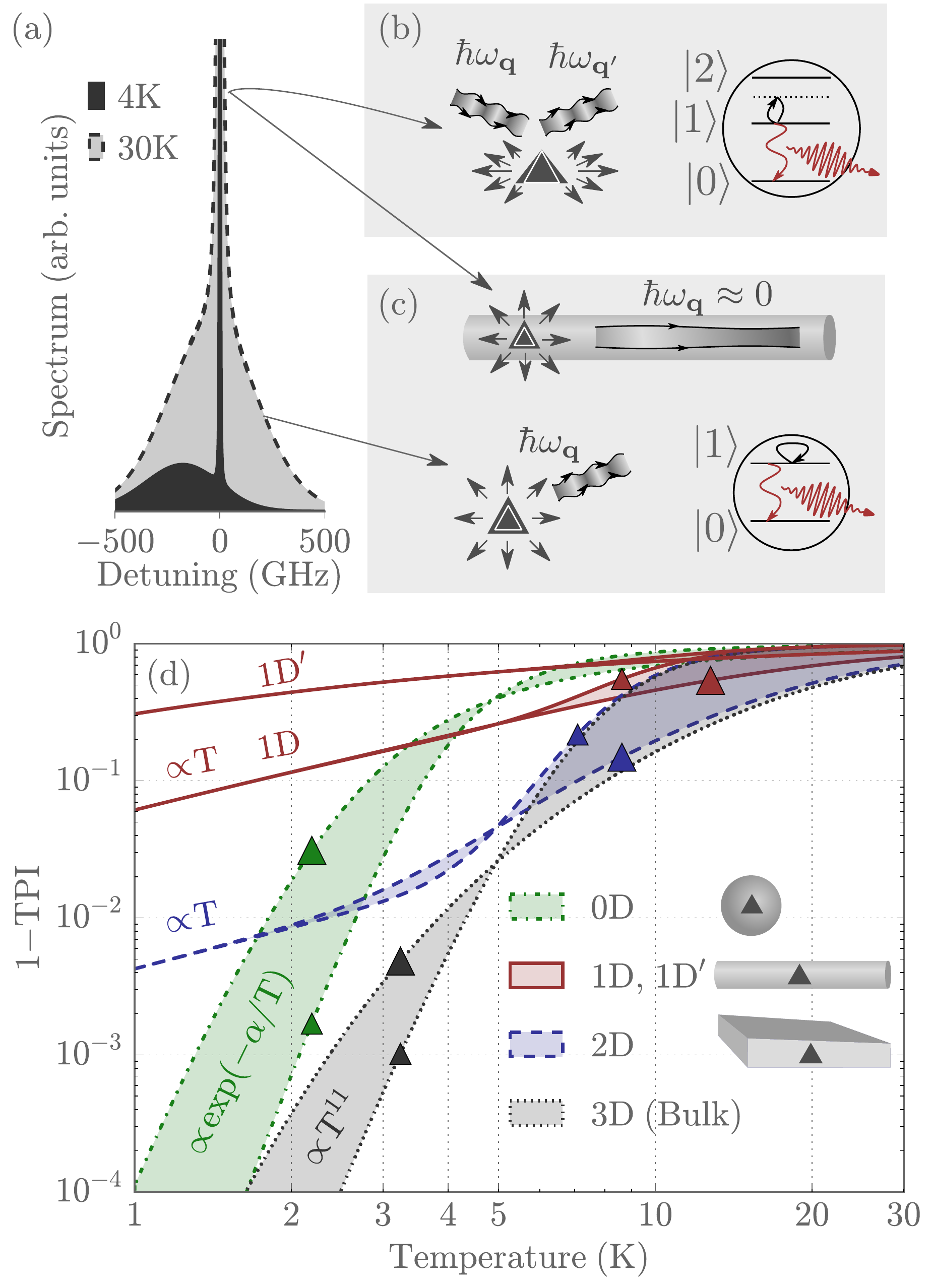}
\caption{\label{Figure1} Phonon dephasing of spontaneous emission from QDs. 
(a) The emission spectrum consists of a ZPL and broad sidebands. 
(b) The quadratic coupling represents scattering of phonons through virtual excitations to a higher state and leads to ZPL broadening. 
(c) The linear coupling is associated with the emission or absorption of phonons by the QD. In bulk this leads to phonon sidebands in the emission. In nano-structures, an additional mechanism broadens the ZPL through long-wavelength deformations. 
(d) Error in two-photon interference versus temperature for QDs embedded in structures with different dimensionality. 
0D corresponds to a QD in the center of a sphere with radius $R=\SI{80}{\nano\meter}$, 1D to a cylindrical waveguide with radius $\rho=\SI{80}{\nano\meter}$ and the QD placed in the cross-sectional center ($\mathrm{1D}$) or halfway offcenter ($\mathrm{1D'}$), 2D to a QD in the center of a freestanding membrane with height $2h=\SI{160}{\nano\meter}$, and 3D to a bulk medium. 
Each structure is represented by two curves that correspond to a large ($L=\SI{4.5}{\nano\meter}$) and small ($L=\SI{1.5}{\nano\meter}$) wave function denoted with large and small triangles, respectively.}
\end{figure}

Here we present a general microscopic theory describing the influence of phonons on the coherence of QDs and the indistinguishability of the emitted photons in photonic (nano)structures. 
The model is applied to the four experimentally relevant systems of an In(Ga)As QD in a cavity, waveguide, slab, or bulk medium corresponding to different geometric dimensionality from 0D to 3D, see Fig.~\ref{Figure1}(d).
The interaction with light is treated phenomenologically as a Markovian decay channel, which is the most abundant situation for applications such as coherent single-photon sources. 
The interplay between coherence and efficiency in the opposite limit of bulk phonons and non-Markovian light-matter interaction was recently explored in Ref.~\cite{IlesSmith17}.

The impact of phonons on the photon emission from QDs can be seen clearly in the emission spectrum~\cite{muljarov04, thoma16, zajac16, muljarov05, kaer12, madsen13, nazir16, borri05, bayer02, langbein04, roy16, roy15, kaer13, grange09, lindwall07, rudin06, besombes01}, which features broad sidebands superimposed on a narrow zero-phonon line (ZPL), cf. Fig.~\ref{Figure1}(a). 
The sidebands originate from rapid phonon emission or absorption on a pico-second time scale, see Fig.~\ref{Figure1}(c), while the ZPL arises from the long-time decay of coherence over nano-second time scales. 
Due to the large spectral mismatch between the two processes, the incoherent sidebands can readily be filtered while maintaining a high efficiency of the photon source~\cite{IlesSmith17,somaschi16,SI_PRL}. 
The fundamental limit to photon indistinguishability is therefore the interaction between the QD and phonons over long time scales, which is the main focus of the present Letter.

In a bulk medium, the broadening of the ZPL is described by an exciton-phonon coupling that is quadratic in phonon displacement~\cite{muljarov04}, cf.~Fig.~\ref{Figure1}(b). 
Here we obtain a simple expression for the dephasing rate, $\Gamma_\mathrm{3D}$,
\begin{equation}
\Gamma_\textrm{3D} = 3\pi \frac{v_s}{L} C_\mathrm{Q}^2\int_0^\infty \mathrm{d}(qL) (qL)^{10} \textrm{e}^{-(qL)^2} N_q (N_q+1),
\label{eq:Gamma_ph_bulk}
\end{equation}
where $C_\mathrm{Q}$ is a dimensionless constant defined later, $L$ the radius of the QD wave function, $v_s$ the speed of sound, and $q$ and $N_q$ the phonon wavenumber and occupation number, respectively. 
Remarkably, when the thermal wavelength is larger than the QD size, $\lambda_\mathrm{th} > L$, corresponding to a temperature below a critical temperature $T_\mathrm{c}=\hbar v_s/k_\mathrm{B}L$, the phonons freeze out leading to a rapid drop of the dephasing rate, cf.~Fig.~\ref{Figure1}(d). 
This yields $\Gamma_\mathrm{3D}(T<T_c)\simeq 3\pi(v_s/L)\times 10!\times C_\mathrm{Q}^2(T/T_\mathrm{c})^{11}$ leading to highly coherent processes at $T\lesssim \SI{4}{\kelvin}$ for realistic QD sizes. 
Nano-structures on the other hand are finite and can thus expand freely resulting in long-wavelength vibrations that broaden the ZPL already within the linear exciton-phonon coupling, see Fig.~\ref{Figure1}(c). 
The latter competes with the quadratic coupling to yield a non-trivial temperature dependence of the photon indistinguishability, cf.~Fig.~\ref{Figure1}(d). 
We find that these processes severely limit the coherence in 1D and 2D nano-structures.

To derive the results we generalize the formalism developed in Ref.~\cite{muljarov04} and consider arbitrary structures with the electron-phonon Hamiltonian
\begin{equation}
\begin{split}
H = \hbar\omega_{01}\ket{1}\bra{1} + \sum_\mathbf{q}\hbar\omega_\qq \ad_\qq a_\qq + V\ket{1}\bra{1},
\end{split}
\end{equation}
where $\ket{1}$ is the QD excited state, $\ad_\qq$ ($a_\qq$) the creation (annihilation) operator for the phonon mode with momentum $\qq$ and energy $\hbar\omega_\qq$, and $\hbar\omega_{01}$ is the QD transition energy, cf.~Fig.~\ref{Figure1}(b). 
We assume low temperatures such that the QD excited states are not populated. 
The interaction term, $V=V_\mathrm{L}+V_\mathrm{Q}$, comprises a linear and a quadratic term in phonon displacement
\begin{equation}
\begin{split}
V_\mathrm{L} &= \sum_\qq L_\qq A_\qq,\hspace{0.3cm} V_\mathrm{Q} = \sum_{b,m}\left[\sum_{\qq}Q_{\qq b}^{m}A_\qq\right]^2,\\
L_\qq &= M_{\qq e}^{11} - M_{\qq h}^{11}, \hspace{1cm} Q_{\qq b}^{m} = \frac{M_{\qq b}^{1m}}{\sqrt{\Delta_m}},
\end{split}
\label{eq:V_L_V_Q}
\end{equation}
where $A_\qq=a_\qq+\ad_\qq$, $b = \{\mathrm{e},\mathrm{h}\}$ denotes electron or hole, $M_{\qq b}^{mn}$ is the electron-phonon matrix element, and $\Delta_m$ is the energy distance between the ground, $\ket{1}$, and $m$-th state of the QD with $m \geq 2$.
Time-reversal symmetry implies that all quantities can be chosen real. 
The interaction with phonons is dominated by the deformation-potential coupling~\cite{SI_PRL,takagahara99}
\begin{equation}
M_{\qq b}^{mn} = D_b \me{\psi_b^{m}}{\divv \mathbf{u}_\qq}{\psi_{b}^{n}},
\label{eq:matrix_element}
\end{equation}
where $D_b$ is the deformation-potential constant, $\psi_b^m$ the wavefunction of the $m$-th state, and $\mathbf{u}_\qq$ the phonon displacement.

After excitation at $t=0$, the QD coherence is described by the correlation function $P(t)=\left< \smm(t)\spp(0) \right>= \left< \mathcal{T}\mathrm{e}^{ -\frac{\mathrm{i}}{\hbar}\int_0^t\mathrm{d}\tau \tilde{V}(\tau) } \right>$~\cite{mahan13,muljarov04},
where $\mathcal{T}$ is the time-ordering operator, and $\tilde{V}$ the potential in the interaction picture with respect to the free phonon Hamiltonian.
$P(t)$ can be evaluated numerically exact using the cumulant expansion~\cite{kubo62,muljarov04} but we follow a simplified approach that captures the essential physics. 
We find that the QD-phonon interaction is weak, such that retaining the first two terms in the cumulant expansion is sufficient. 
The distribution therefore becomes Gaussian and is completely specified by the mean and standard deviation of the noise $F(t) = -(\mathrm{i}/\hbar)\int_0^t\mathrm{d}\tau \tilde{V}(\tau)$. 
The truncation is thus equivalent to treating the phonon bath as a Gaussian noise source, which yields $P(t) \simeq \mathrm{exp}(-\mathrm{i}\mu_F)\exp\left[ -\frac{1}{2}\left( \av{\mathcal{T}F^2(t)} - \mu_F^2 \right)\right]$, where $\mu_F = \av{F(t)}$. 
In~\cite{SI_PRL} we show that this provides an excellent approximation to the exact numerical result. 
Inserting \eqref{eq:V_L_V_Q} into the above expression yields
\begin{equation}
\begin{split}
P(t) &= \exp\left[ -\mathrm{i}\mu_F + K_\mathrm{L}(t) + K_\mathrm{Q}(t) \right],\\
K_\mathrm{L}(t) &= -\frac{\mathrm{i}}{2\hbar}\sum_{\qq}\abs{L_\qq}^2\iint_0^t\mathrm{d}t\mathrm{d}t' D_\qq(t-t'),\\
K_\mathrm{Q}(t) &= \sum_{b mn}\iint_0^t\mathrm{d}t\mathrm{d}t' \left[\sum_\qq Q_{\qq b}^{m}Q_{\qq b}^{n} D_\qq(t-t')\right]^2,
\end{split}
\label{eq:K_L_K_Q}
\end{equation}
where $D_\qq(t) = (-\mathrm{i}/\hbar)\left[ \left(N_\qq+1\right)\textrm{e}^{-\mathrm{i}\omega_\qq \abs{t}} + N_\qq \textrm{e}^{\mathrm{i}\omega_\qq \abs{t}} \right]$ is the phonon Green function. The function $K_\mathrm{L}(t)$ stems from the linear electron-phonon interaction and is determined by a matrix element of the form $M_{\qq b}^{11}\propto \me{\psi_b^1}{\divv \mathbf{u}_\qq}{\psi_b^1}$, which shares the symmetry of the ground-state wave function, implying that $K_\mathrm{L}(t)$ couples to symmetric acoustic deformations, cf.~Fig.~\ref{Figure1}(c). On the other hand, $K_\mathrm{Q}(t)$ is mediated by phonons that share the symmetry of the excited states. 

\begin{figure}
\includegraphics[width=\columnwidth]{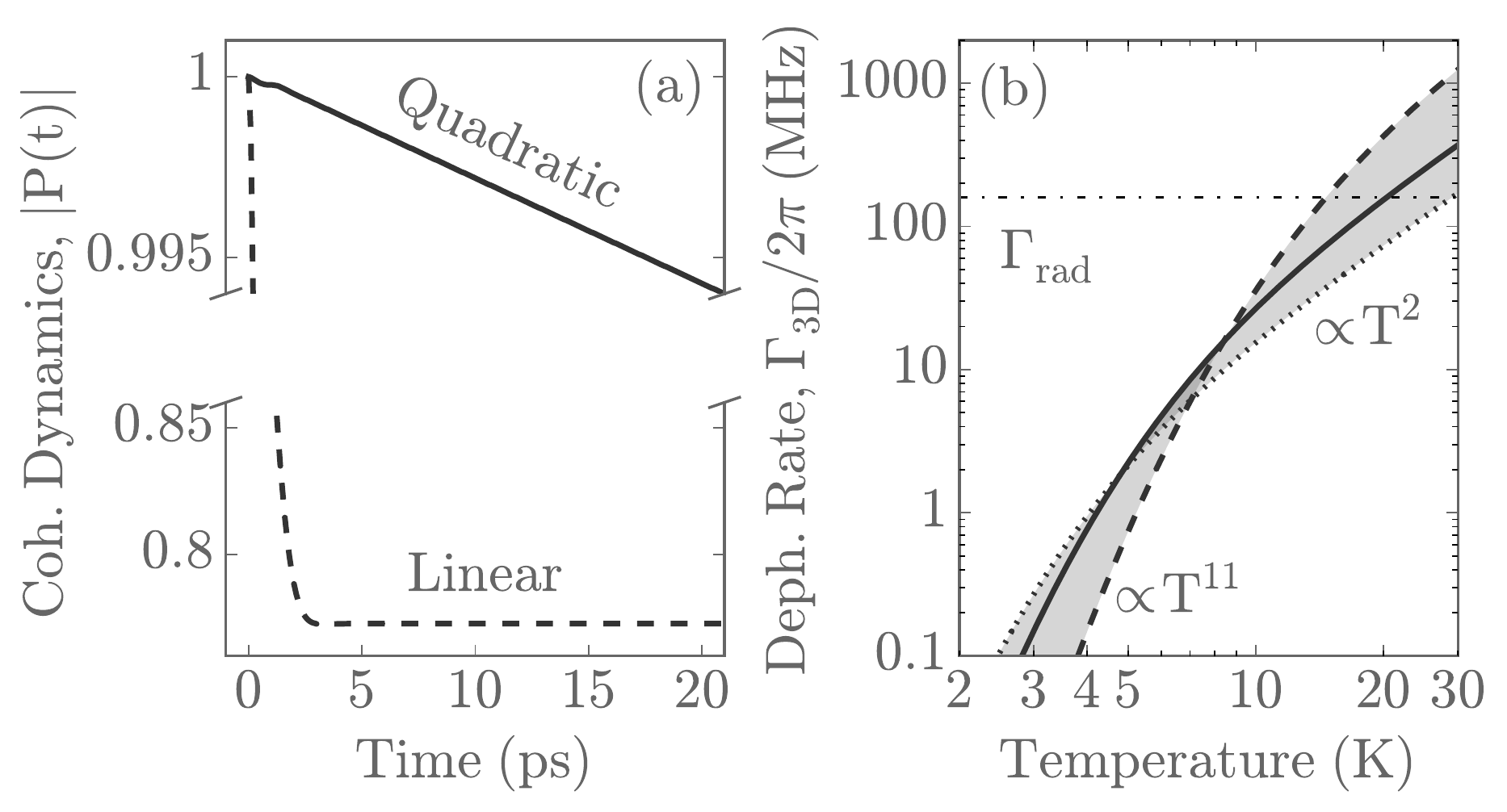}
\caption{\label{Figure2} Phonon dephasing in a bulk medium. (a) The linear (quadratic) exciton-phonon coupling affects the short-time (long-time) decay of coherence. Parameters: $T=\SI{10}{\kelvin}$, $L=\SI{3}{\nano\meter}$. (b) Phonon dephasing rate versus temperature for $L=\SI{1.5}{\nano\meter}$ (dashed line), $L=\SI{3}{\nano\meter}$ (solid line) and $L=\SI{4.5}{\nano\meter}$ (dotted line). The natural linewidth in a bulk medium is indicated by the dash-dotted line.
}
\end{figure}

In a bulk medium, the linear interaction $K_\mathrm{L}(t)$ does not contribute to the long-time decay of coherence, see Fig.~\ref{Figure2}(a). 
The quadratic coherence function $P_\mathrm{Q}(t)=\exp[K_\mathrm{Q}(t)]$ is evaluated for a spherical QD with Gaussian envelopes of radius $L$. 
Since $K_\mathrm{Q}(t)$ is proportional to $\Delta_m^{-2}$, the inclusion of the first triply degenerate excited state, $m=n=2$, gives the dominant contribution. 
Evaluating $P_\mathrm{Q}(t)$ numerically yields a Markovian decay over long time scales, cf.~Fig.~\ref{Figure2}(a), with $\mathrm{Re}\left[K_\mathrm{Q}(t)\right] = -\Gamma_\mathrm{3D}t$, and $\mathrm{Im}\left[K_\mathrm{Q}(t)\right]$ contributes to a spectral shift. 
The dephasing rate can be calculated analytically by performing the time integration in \eqref{eq:K_L_K_Q} and using the long-time limit $\omega_\qq^{-2}\sin^2\omega_\qq t \simeq \pi t\delta(\omega_\qq)$. 
This leads to \eqref{eq:Gamma_ph_bulk} with $C_\mathrm{Q} = \left( D_e^2/\Delta_e + D_h^2/\Delta_h \right)/3 (2\pi)^2\rho_m v_s^2 L^3$. This is plotted in Fig.~\ref{Figure2}(b) for GaAs parameters: $v_s=\SI{4780}{\meter\second^{-1}}$, mass density $\rho_m=\SI{5.37}{\gram\centi\meter^{-3}}$, $D_\mathrm{e} = \SI{-14.6}{\electronvolt}$ and $D_\mathrm{h} = \SI{-4.8}{\electronvolt}$. 
The energy distance to the excited states is taken to be $\Delta_e=2\Delta_h=\SI{40}{\milli\electronvolt}\times L_0/L$ with $L_0=\SI{3}{\nano\meter}$, in accordance with theoretical estimates and experimental results~\cite{schwartz16, pryor98, schmidt96, noda98, grundmann95, landin98, stier99}.
This choice of parameters is justified in~\cite{SI_PRL}.

To relate the phonon decoherence to the quality of the photons emitted by the QD, we study a Hong-Ou-Mandel setup~\cite{santori02}.  Here, the second-order correlation function determines the two-photon indistinguishability (TPI), which ranges from 0 (no indistinguishability) to 1 (perfect indistinguishability). 
If (i) the QD-light interaction is Markovian, (ii) the excitation happens instantaneously, (iii) the QD is a perfectly antibunched source of single photons, and (iv) the noise is stationary, the TPI is~\cite{kiraz04}
\begin{equation}
\mathrm{TPI} = \Gamma_\mathrm{rad}\int_0^\infty \mathrm{d}\tau\mathrm{e}^{-\Gamma_\mathrm{rad}\tau}\abs{P(\tau)}^2.
\label{eq:TPI}
\end{equation}
In bulk, $\abs{P(t)}\simeq\exp(-\Gamma_\mathrm{3D}t)$ leading to $\mathrm{TPI} = \Gamma_\mathrm{rad}/(\Gamma_\mathrm{rad}+2\Gamma_\mathrm{3D})$ after filtering out the sidebands, where $\Gamma_\textrm{rad}\simeq 2\pi\times \SI{160}{MHz}$ is the radiative decay rate of the QD~\cite{johansen08}. 
The resulting temperature dependence of the $\mathrm{TPI}$ is plotted in Fig.~\ref{Figure1}(d). Near-unity indistinguishabilities can be achieved at temperatures below a few Kelvin. 
Analytic solutions can also be found at high temperatures, $\Gamma_\mathrm{3D}(T>T_c)\simeq 3\pi^{3/2}\times (105/32)(v_s/L)\times C_\mathrm{Q}^2(T/T_c)^2$, with a quadratic temperature dependence.

In the following we study phonon decoherence in nano-structures~\cite{arcari14,gazzano13,madsen14,bennett16}. 
The short-time dynamics results in phonon sidebands that are shaped by the density of states, but this modification is not significant for the sizes considered here~\cite{SI_PRL}.
The long-time dynamics can be split into two contributions, $P_\mathrm{ZPL}(t)=P_\mathrm{Q}(t)  P_\mathrm{L0}(t)$,
where $P_\mathrm{Q}(t)$ stems from the quadratic coupling, and $P_\mathrm{L0}(t)$ is a nano-structure-specific low-frequency contribution to the linear coupling ~\cite{palma96, lindwall07}.
For simplicity we keep a fixed bulk-like radiative decay rate throughout this work. 
In a realistic device this value may differ in which case the results should be modified accordingly.

\begin{figure}
\includegraphics[width=\columnwidth]{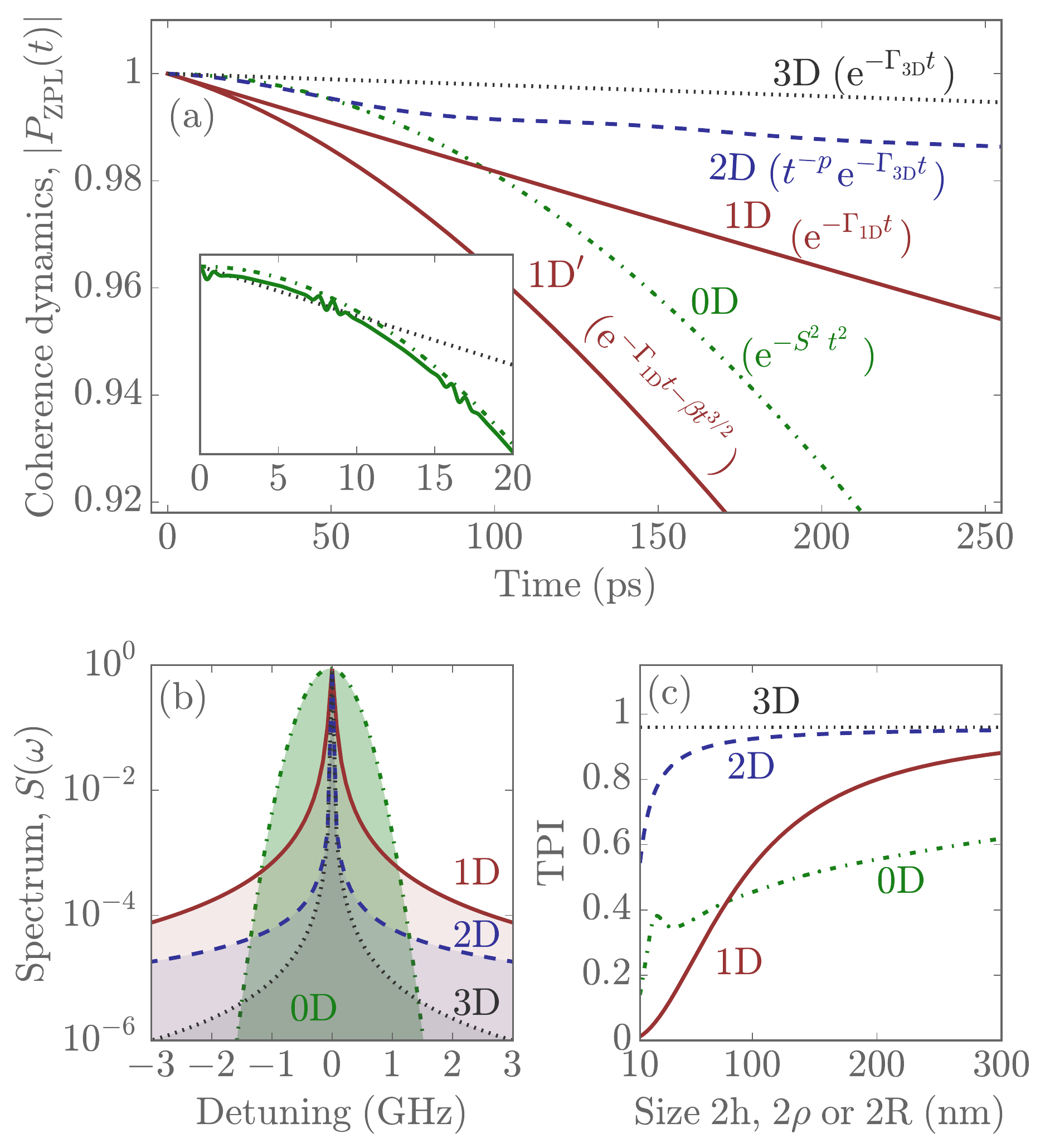}
\caption{\label{Figure3} Phonon dephasing in photonic structures and the role of dimensionality. (a) Decay of coherence for the same structures as in Fig.~\ref{Figure1}(d) except the 0D structure ($R=\SI{20}{\nano\meter}$). Inset: initial decay of coherence (0--\SI{20}{\pico\second}) for bulk (dotted line), sphere (solid line), and the $t^2$-approximation for the sphere. (b) Corresponding emission spectrum. (c) Two-photon indistinguishability versus size of the structure. All plots are for $T=\SI{5}{\kelvin}$, $L=\SI{3}{\nano\meter}$.
}
\end{figure}

We start with a 0D nano-sphere cavity, which resembles the geometry of colloidal QDs embedded in spherical shells~\cite{werschler16}. 
The long-time coherence is plotted in Fig.~\ref{Figure3}(a) and stems solely from the quadratic coupling $P_\mathrm{Q}(t)$. 
The decay is strongly non-Markovian because the phonons are reflected at the boundary and interact with the QD periodically, see the inset of Fig.~\ref{Figure3}(a). 
A simple expression for $\mathrm{Re}\left[ K_\mathrm{Q}(t)\right]$ can be derived from \eqref{eq:K_L_K_Q} by using the long-time form $\sin^2[(\omega_{j}-\omega_{j'}) t]/(\omega_j-\omega_{j'})^2\simeq t^2\delta_{jj'}$, where $j$ is the index of the confined acoustic mode. 
This results in $P_\mathrm{ZPL}(t) = \exp(-S^2t^2)$ with
\begin{equation}
S^2 = \frac{3}{2}\left(\frac{\pi}{2}\frac{v_s}{L}C_\mathrm{Q}\right)^2\sum_j I_{j}^4 \tilde{q}_j^{12} \mathrm{e}^{-\tilde{q}_j^2}N_{\tilde{q}_j}\left( N_{\tilde{q}_j}+1 \right),
\end{equation}
where $\tilde{q}_j\equiv q_j L$, and $I_j$ is a dimensionless normalization factor of the ($j$,1,0) spheroidal mode~\cite{masumoto13,SI_PRL}. 
The resulting emission spectrum without the radiative broadening, $S(\omega) = \mathrm{Re}\int_0^\infty\mathrm{d}tP_\mathrm{ZPL}(t)\exp(-\mathrm{i}\omega t)$, is a Gaussian as depicted in Fig.~\ref{Figure3}(b). 
The TPI yields $\mathrm{TPI}_\mathrm{0D} = \sqrt{\pi}r_\mathrm{s}\exp(r_\mathrm{s}^2)\mathrm{erfc}(r_\mathrm{s})$,
where $r_\mathrm{s} = \Gamma_\mathrm{rad}/2\sqrt{2}S$, and is plotted in Fig.~\ref{Figure1}(d). 
In general, the decoherence is stronger than in bulk. 
However, in the small temperature limit $\lambda_\mathrm{th}\gg R$, the thermal energy is smaller than the lowest vibrational state of the sphere leading to negligible decoherence, $1-\mathrm{TPI}_\mathrm{0D}\propto \exp(-\hbar\omega/2k_\mathrm{B}T)$, as depicted in Fig.~\ref{Figure1}(d).

In the following we discuss the dephasing of QDs embedded in 1D and 2D structures~\cite{coles16,heiss13,arcari14}. 
The quadratic interaction does not deviate significantly from bulk because $K_\mathrm{Q}(t)$ is dominated by phonons with a wavelength comparable to the QD size while realistic photonic structures are much larger and do not affect those phonon modes. 
We therefore assume $K_\mathrm{Q}^\mathrm{1D}\simeq K_\mathrm{Q}^\mathrm{2D} \simeq -\Gamma_\mathrm{3D}t$~\cite{SI_PRL}. 
This is different for the linear interaction $P_\mathrm{L0}$~\cite{lindwall07,galland08}. 
For a freestanding 1D waveguide two families of acoustic modes contribute to dephasing with a finite $\divv \mathbf{u}$: longitudinal expansions of the rod with a linear dispersion and thus a constant density of states at $\omega \rightarrow 0$, and flexural modes that bend the rod with a quadratic dispersion and a diverging density of states~\cite{SI_PRL}. 
The former yield a Markovian decay that was found in Ref.~\cite{lindwall07} for a cylinder but here is generalized to an arbitrary cross-sectional shape and QD position with the rate 
\begin{equation}
\Gamma_\mathrm{L0}^\mathrm{1D} = \frac{(D_e-D_h)^2(1-2\nu)^2k_\mathrm{B}T}{2A\rho_mv_\mathrm{1D}^3\hbar^2},
\label{eq:K1D}
\end{equation}
where $A$ is the cross-sectional area, $\nu=0.299$ the GaAs Poisson ratio, and $v_\mathrm{1D}=v_s\sqrt{3+2\nu+2/(\nu-1)}$ the phonon speed along the waveguide axis. 
The total decay, $\exp(-\Gamma_\mathrm{1D}t)$, with $\Gamma_\mathrm{1D}=\Gamma_\mathrm{L0}^\mathrm{1D} + \Gamma_\mathrm{3D}$, is plotted in Fig.~\ref{Figure3}(a). 
The coupling to flexural modes, on the other hand, depends on the QD position, and ranges from no coupling at points of high symmetry (e.g., the center of a cylinder) to large coupling away from such points. 
In Fig.~\ref{Figure3}(a) we plot the numerically evaluated coherence decay of a QD placed offcenter at a distance $\rho/2$ from the center of a cylindrical waveguide of radius $\rho$ (1D'). 
The decay is non-Markovian scaling as $P_\mathrm{1D'}=\mathrm{e}^{-\Gamma_\mathrm{1D}t-\beta t^{3/2}}$~\cite{SI_PRL}. 
In both cases, the error in TPI scales as $\propto T$ at low temperatures and is significant even for a waveguide with a diameter of hundreds of nanometers, see Fig.~\ref{Figure1}(d) and Fig.~\ref{Figure3}(c).

Next we consider a QD embedded in a freestanding 2D membrane with thickness $2h$. 
To evaluate the coherence, we approximate the dispersion of the fundamental vibrational mode~\cite{anghel07} as $\omega=v_\mathrm{2D}q_{||}$ and discard the modes with $q_{||} > h^{-1}$~\cite{SI_PRL}, where $v_\mathrm{2D}=v_s\sqrt{1-2\nu}/(1-\nu)$, and $q_{||}$ is the in-plane wave number. 
The linear scaling of the density of states with $\omega$ yields
\begin{equation}
\begin{split}
\mathrm{Re}\left[K_\mathrm{2D}\right] = -p \left[ \gamma_\mathrm{E}+\int_{\tilde{t}}^\infty\mathrm{d}\tau\frac{\cos\tau}{\tau}+\ln \tilde{t} \right]-\Gamma_\mathrm{3D}t,
\end{split}
\label{eq:K2D}
\end{equation}
Here, $p=(D_e-D_h)^2(1-2\nu)^2k_\mathrm{B}T/4\pi \rho_m h v_\mathrm{2D}^4(1-\nu)^2\hbar^2$, $\tilde{t} = v_\mathrm{2D}t/h$, and $\gamma_\mathrm{E}$ is the Euler-Mascheroni constant. 
The coherence is plotted in Fig.~\ref{Figure3}(a) and results in a TPI that is dominated by the linear interaction at low temperatures as shown in Fig.~\ref{Figure1}(d). 
At long times ($t\gg h/v_\mathrm{2D}$), \eqref{eq:K2D} can be simplified to $P_\mathrm{2D}=(v_\mathrm{2D}t/h)^{-p}\exp(-\Gamma_\mathrm{3D}t)$. 
QDs positioned away from the membrane center would also couple to flexural modes with quadratic dispersion resulting in a Markovian dephasing. 
Contrary to these examples, the 0D structure has a vanishing density of states at low frequencies. 
As used above, the dephasing is therefore only due to the quadratic coupling.

\begin{figure}
\includegraphics[width=0.75\columnwidth]{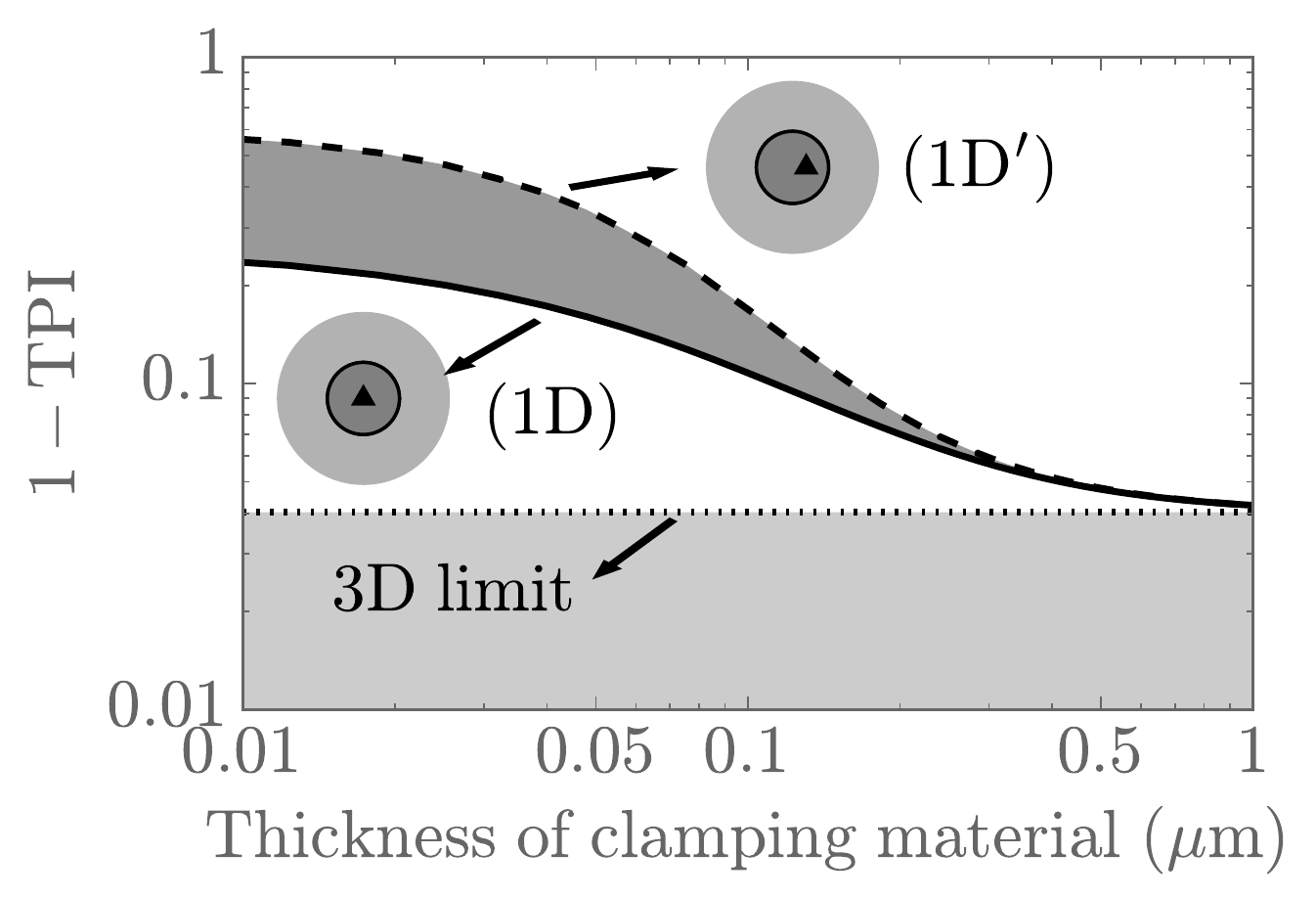}
\caption{\label{Figure4} Suppressing phonon dephasing by clamping the photonic structures. The plot shows the photon infidelity versus thickness of SiO$_2$ (gray) surrounding a GaAs waveguide (dark gray) of radius \SI{80}{\nano\meter} with a QD (small black triangle) in the center (solid line) and halfway offcenter (dashed line). Parameters: $T=\SI{5}{\kelvin}$, $L = \SI{3}{\nano\meter}$, $v_{s,\mathrm{SiO2}} = \SI{5848}{\meter\second^{-1}}$, $\rho_{m,\mathrm{SiO2}} = \SI{2.2}{\gram\centi\meter^{-3}}$, $\nu_{\mathrm{SiO2}} = 0.17$.
}
\end{figure}

The theory developed above directly points towards methods of suppressing the malign impact of phonons. 
By mechanically clamping the structure, the coupling to the fundamental vibrational mode can be suppressed. 
This may be achieved by immersing the freestanding structure into another material (e.g., glass or semiconductor~\cite{midolo12}) with a lower refractive index such that the light is still guided. 
We find that thicknesses as small as one \si{\micro\meter} are sufficient to fully suppress the decoherence, see Fig.~\ref{Figure4}.
This may provide a viable approach to obtain near-unity TPI.

In conclusion, we find that the degree of confinement of the nano-structure has a significant impact on the coherence. 
Bulk (3D) and maximally confined (0D) structures dephase the ZPL solely due to the quadratic exciton-phonon coupling, which becomes negligible at low temperatures, specifically for $\lambda_\mathrm{th}\gg L$ and $\lambda_\mathrm{th}\gg R$, respectively. 
The decoherence in 1D and 2D structures is enhanced by long-wavelength vibrations mediated by the linear exciton-phonon coupling and can be important even at sub-Kelvin temperatures.
The situation is more involved in the case of more complex structures such as photonic-crystal devices.
We expect a photonic-crystal membrane to exhibit worse coherence than a 2D membrane of same thickness due to the holes that would slow down the long-wavelength phonons, see \eqref{eq:K2D}.
A detailed calculation of the decoherence in photonic-crystal structures is an interesting question for further investigation.

\begin{acknowledgments}
During the final stages of this work a related preprint appeared, which studies the phonon decoherence in bulk systems~\cite{reigue16}. 
We thank Johan R. Ott and Anna Grodecka-Grad for their collaboration at an early stage of this work, and Jake Iles-Smith and Jesper M\o rk for useful discussions. 
We gratefully acknowledge the financial support from the Danish Council for Independent Research, the European Research Council (ERC consolidator grant "QIOS" and advanced grant "SCALE"), and the Academy of Finland through its Centre of Excellence program (project no. 312299).
\end{acknowledgments}

\bibliographystyle{naturemag}
\bibliography{PhononDecoherenceQDs}

\begin{thebibliography}{10}
\expandafter\ifx\csname url\endcsname\relax
  \def\url#1{\texttt{#1}}\fi
\expandafter\ifx\csname urlprefix\endcsname\relax\def\urlprefix{URL }\fi
\providecommand{\bibinfo}[2]{#2}
\providecommand{\eprint}[2][]{\url{#2}}

\bibitem{lodahl15}
\bibinfo{author}{Lodahl, P.}, \bibinfo{author}{Mahmoodian, S.} \&
  \bibinfo{author}{Stobbe, S.}
\newblock \bibinfo{title}{Interfacing single photons and single quantum dots
  with photonic nanostructures}.
\newblock \emph{\bibinfo{journal}{Rev. Mod. Phys.}}
  \textbf{\bibinfo{volume}{87}}, \bibinfo{pages}{347} (\bibinfo{year}{2015}).
\newblock \urlprefix\url{http://dx.doi.org/10.1103/RevModPhys.87.347}.

\bibitem{kuhlmann13}
\bibinfo{author}{Kuhlmann, A.~V.} \emph{et~al.}
\newblock \bibinfo{title}{{Charge noise and spin noise in a semiconductor
  quantum device}}.
\newblock \emph{\bibinfo{journal}{Nat. Phys.}} \textbf{\bibinfo{volume}{9}},
  \bibinfo{pages}{570} (\bibinfo{year}{2013}).

\bibitem{arnold14}
\bibinfo{author}{Arnold, C.} \emph{et~al.}
\newblock \bibinfo{title}{{Cavity-Enhanced Real-Time Monitoring of
  Single-Charge Jumps at the Microsecond Time Scale}}.
\newblock \emph{\bibinfo{journal}{Phys. Rev. X}} \textbf{\bibinfo{volume}{4}},
  \bibinfo{pages}{021004} (\bibinfo{year}{2014}).
\newblock \urlprefix\url{http://dx.doi.org/10.1103/PhysRevX.4.021004}.

\bibitem{urbaszek13}
\bibinfo{author}{Urbaszek, B.} \emph{et~al.}
\newblock \bibinfo{title}{{Nuclear spin physics in quantum dots: An optical
  investigation}}.
\newblock \emph{\bibinfo{journal}{Rev. Mod. Phys.}}
  \textbf{\bibinfo{volume}{85}}, \bibinfo{pages}{79} (\bibinfo{year}{2013}).
\newblock \urlprefix\url{http://dx.doi.org/10.1103/RevModPhys.85.79}.

\bibitem{delteil14}
\bibinfo{author}{Delteil, A.}, \bibinfo{author}{Gao, W.-b.},
  \bibinfo{author}{Fallahi, P.}, \bibinfo{author}{Miguel-Sanchez, J.} \&
  \bibinfo{author}{Imamo\ifmmode~\breve{g}\else \u{g}\fi{}lu, A.}
\newblock \bibinfo{title}{{Observation of Quantum Jumps of a Single Quantum Dot
  Spin Using Submicrosecond Single-Shot Optical Readout}}.
\newblock \emph{\bibinfo{journal}{Phys. Rev. Lett.}}
  \textbf{\bibinfo{volume}{112}}, \bibinfo{pages}{116802}
  (\bibinfo{year}{2014}).
\newblock \urlprefix\url{http://dx.doi.org/10.1103/PhysRevLett.112.116802}.

\bibitem{kuhlmann15}
\bibinfo{author}{Kuhlmann, A.~V.} \emph{et~al.}
\newblock \bibinfo{title}{{Transform-limited single photons from a single
  quantum dot}}.
\newblock \emph{\bibinfo{journal}{Nat. Commun.}} \textbf{\bibinfo{volume}{6}},
  \bibinfo{pages}{8204} (\bibinfo{year}{2015}).
\newblock \urlprefix\url{http://dx.doi.org/10.1038/ncomms9204}.

\bibitem{somaschi16}
\bibinfo{author}{Somaschi, N.} \emph{et~al.}
\newblock \bibinfo{title}{{Near-optimal single-photon sources in the solid
  state}}.
\newblock \emph{\bibinfo{journal}{Nat. Photon.}} \textbf{\bibinfo{volume}{10}},
  \bibinfo{pages}{340} (\bibinfo{year}{2016}).
\newblock \urlprefix\url{http://dx.doi.org/10.1038/nphoton.2016.23}.

\bibitem{ding16}
\bibinfo{author}{Ding, X.} \emph{et~al.}
\newblock \bibinfo{title}{On-demand single photons with high extraction
  efficiency and near-unity indistinguishability from a resonantly driven
  quantum dot in a micropillar}.
\newblock \emph{\bibinfo{journal}{Phys. Rev. Lett.}}
  \textbf{\bibinfo{volume}{116}}, \bibinfo{pages}{020401}
  (\bibinfo{year}{2016}).
\newblock \urlprefix\url{http://dx.doi.org/10.1103/PhysRevLett.116.020401}.

\bibitem{he13}
\bibinfo{author}{He, Y.-M.} \emph{et~al.}
\newblock \bibinfo{title}{{On-demand semiconductor single-photon source with
  near-unity indistinguishability}}.
\newblock \emph{\bibinfo{journal}{Nat. Nanotechnol.}}
  \textbf{\bibinfo{volume}{8}}, \bibinfo{pages}{213} (\bibinfo{year}{2013}).
\newblock \urlprefix\url{http://dx.doi.org/10.1038/nnano.2012.262}.

\bibitem{majumdar11}
\bibinfo{author}{Majumdar, A.}, \bibinfo{author}{Kim, E.~D.} \&
  \bibinfo{author}{Vu\ifmmode \check{c}\else
  \v{c}\fi{}kovi\ifmmode~\acute{c}\else \'{c}\fi{}, J.}
\newblock \bibinfo{title}{Effect of photogenerated carriers on the spectral
  diffusion of a quantum dot coupled to a photonic crystal cavity}.
\newblock \emph{\bibinfo{journal}{Phys. Rev. B}} \textbf{\bibinfo{volume}{84}},
  \bibinfo{pages}{195304} (\bibinfo{year}{2011}).
\newblock \urlprefix\url{https://link.aps.org/doi/10.1103/PhysRevB.84.195304}.

\bibitem{kirsanske17}
\bibinfo{author}{Kir\ifmmode \check{s}\else \v{s}\fi{}ansk\ifmmode~\dot{e}\else
  \.{e}\fi{}, G.} \emph{et~al.}
\newblock \bibinfo{title}{Indistinguishable and efficient single photons from a
  quantum dot in a planar nanobeam waveguide}.
\newblock \emph{\bibinfo{journal}{Phys. Rev. B}} \textbf{\bibinfo{volume}{96}},
  \bibinfo{pages}{165306} (\bibinfo{year}{2017}).
\newblock \urlprefix\url{https://link.aps.org/doi/10.1103/PhysRevB.96.165306}.

\bibitem{thyrrestrup18}
\bibinfo{author}{Thyrrestrup, H.} \emph{et~al.}
\newblock \bibinfo{title}{Quantum optics with near-lifetime-limited quantum-dot
  transitions in a nanophotonic waveguide}.
\newblock \emph{\bibinfo{journal}{Nano Letters}} \textbf{\bibinfo{volume}{18}},
  \bibinfo{pages}{1801--1806} (\bibinfo{year}{2018}).
\newblock \urlprefix\url{https://doi.org/10.1021/acs.nanolett.7b05016}.
\newblock \bibinfo{note}{PMID: 29494160},
  \eprint{https://doi.org/10.1021/acs.nanolett.7b05016}.

\bibitem{besombes01}
\bibinfo{author}{Besombes, L.}, \bibinfo{author}{Kheng, K.},
  \bibinfo{author}{Marsal, L.} \& \bibinfo{author}{Mariette, H.}
\newblock \bibinfo{title}{{Acoustic phonon broadening mechanism in single
  quantum dot emission}}.
\newblock \emph{\bibinfo{journal}{Phys. Rev. B}} \textbf{\bibinfo{volume}{63}},
  \bibinfo{pages}{155307} (\bibinfo{year}{2001}).
\newblock \urlprefix\url{http://dx.doi.org/10.1103/PhysRevB.63.155307}.

\bibitem{muljarov04}
\bibinfo{author}{Muljarov, E.~A.} \& \bibinfo{author}{Zimmermann, R.}
\newblock \bibinfo{title}{{Dephasing in Quantum Dots: Quadratic Coupling to
  Acoustic Phonons}}.
\newblock \emph{\bibinfo{journal}{Phys. Rev. Lett.}}
  \textbf{\bibinfo{volume}{93}}, \bibinfo{pages}{237401}
  (\bibinfo{year}{2004}).
\newblock \urlprefix\url{http://dx.doi.org/10.1103/PhysRevLett.93.237401}.

\bibitem{grange09}
\bibinfo{author}{Grange, T.}
\newblock \bibinfo{title}{{Decoherence in quantum dots due to real and virtual
  transitions: A nonperturbative calculation}}.
\newblock \emph{\bibinfo{journal}{Phys. Rev. B}} \textbf{\bibinfo{volume}{80}},
  \bibinfo{pages}{245310} (\bibinfo{year}{2009}).
\newblock \urlprefix\url{http://dx.doi.org/10.1103/PhysRevB.80.245310}.

\bibitem{thoma16}
\bibinfo{author}{Thoma, A.} \emph{et~al.}
\newblock \bibinfo{title}{{Exploring Dephasing of a Solid-State Quantum Emitter
  via Time- and Temperature-Dependent Hong-Ou-Mandel Experiments}}.
\newblock \emph{\bibinfo{journal}{Phys. Rev. Lett.}}
  \textbf{\bibinfo{volume}{116}}, \bibinfo{pages}{033601}
  (\bibinfo{year}{2016}).
\newblock \urlprefix\url{http://dx.doi.org/10.1103/PhysRevLett.116.033601}.

\bibitem{zajac16}
\bibinfo{author}{Zajac, J.~M.} \& \bibinfo{author}{Erlingsson, S.~I.}
\newblock \bibinfo{title}{{Temperature dependency of resonance fluorescence
  from InAs/GaAs quantum dots: Dephasing mechanisms}}.
\newblock \emph{\bibinfo{journal}{Phys. Rev. B}} \textbf{\bibinfo{volume}{94}},
  \bibinfo{pages}{035432} (\bibinfo{year}{2016}).
\newblock \urlprefix\url{http://dx.doi.org/10.1103/PhysRevB.94.035432}.

\bibitem{claudon10}
\bibinfo{author}{Claudon, J.} \emph{et~al.}
\newblock \bibinfo{title}{A highly efficient single-photon source based on a
  quantum dot in a photonic nanowire}.
\newblock \emph{\bibinfo{journal}{Nat Photon}} \textbf{\bibinfo{volume}{4}},
  \bibinfo{pages}{174--177} (\bibinfo{year}{2010}).
\newblock \urlprefix\url{http://dx.doi.org/10.1038/nphoton.2009.287}.

\bibitem{gazzano13}
\bibinfo{author}{Gazzano, O.} \emph{et~al.}
\newblock \bibinfo{title}{Bright solid-state sources of indistinguishable
  single photons}.
\newblock \emph{\bibinfo{journal}{Nat. Commun.}} \textbf{\bibinfo{volume}{4}},
  \bibinfo{pages}{1425} (\bibinfo{year}{2013}).
\newblock \urlprefix\url{http://dx.doi.org/10.1038/ncomms2434}.

\bibitem{arcari14}
\bibinfo{author}{Arcari, M.} \emph{et~al.}
\newblock \bibinfo{title}{Near-unity coupling efficiency of a quantum emitter
  to a photonic crystal waveguide}.
\newblock \emph{\bibinfo{journal}{Phys. Rev. Lett.}}
  \textbf{\bibinfo{volume}{113}}, \bibinfo{pages}{093603}
  (\bibinfo{year}{2014}).
\newblock \urlprefix\url{http://dx.doi.org/10.1103/PhysRevLett.113.093603}.

\bibitem{palma96}
\bibinfo{author}{Palma, G.}, \bibinfo{author}{Suominen, K.-A.} \&
  \bibinfo{author}{Ekert, A.~K.}
\newblock \bibinfo{title}{Quantum computers and dissipation}.
\newblock \emph{\bibinfo{journal}{Proc. R. Soc. Lond. A}}
  \textbf{\bibinfo{volume}{452}}, \bibinfo{pages}{567} (\bibinfo{year}{1996}).
\newblock \urlprefix\url{http://dx.doi.org/10.1098/rspa.1996.0029}.

\bibitem{lindwall07}
\bibinfo{author}{Lindwall, G.}, \bibinfo{author}{Wacker, A.},
  \bibinfo{author}{Weber, C.} \& \bibinfo{author}{Knorr, A.}
\newblock \bibinfo{title}{{Zero-Phonon Linewidth and Phonon Satellites in the
  Optical Absorption of Nanowire-Based Quantum Dots}}.
\newblock \emph{\bibinfo{journal}{Phys. Rev. Lett.}}
  \textbf{\bibinfo{volume}{99}}, \bibinfo{pages}{087401}
  (\bibinfo{year}{2007}).
\newblock \urlprefix\url{http://dx.doi.org/10.1103/PhysRevLett.99.087401}.

\bibitem{galland08}
\bibinfo{author}{Galland, C.}, \bibinfo{author}{H\"ogele, A.},
  \bibinfo{author}{T\"ureci, H.~E.} \&
  \bibinfo{author}{Imamo\ifmmode~\breve{g}\else \u{g}\fi{}lu, A.}
\newblock \bibinfo{title}{Non-markovian decoherence of localized nanotube
  excitons by acoustic phonons}.
\newblock \emph{\bibinfo{journal}{Phys. Rev. Lett.}}
  \textbf{\bibinfo{volume}{101}}, \bibinfo{pages}{067402}
  (\bibinfo{year}{2008}).
\newblock \urlprefix\url{http://dx.doi.org/10.1103/PhysRevLett.101.067402}.

\bibitem{IlesSmith17}
\bibinfo{author}{Iles-Smith, J.}, \bibinfo{author}{McCutcheon, D. P.~S.},
  \bibinfo{author}{Nazir, A.} \& \bibinfo{author}{M\o{}rk, J.}
\newblock \bibinfo{title}{Phonon limit to simultaneous near-unity efficiency
  and indistinguishability in semiconductor single photon sources}.
\newblock \emph{\bibinfo{journal}{Nat. Photon.}} \textbf{\bibinfo{volume}{11}},
  \bibinfo{pages}{521} (\bibinfo{year}{2017}).
\newblock \urlprefix\url{https://dx.doi.org/10.1038/nphoton.2017.101}.

\bibitem{muljarov05}
\bibinfo{author}{Muljarov, E.~A.}, \bibinfo{author}{Takagahara, T.} \&
  \bibinfo{author}{Zimmermann, R.}
\newblock \bibinfo{title}{{Phonon-Induced Exciton Dephasing in Quantum Dot
  Molecules}}.
\newblock \emph{\bibinfo{journal}{Phys. Rev. Lett.}}
  \textbf{\bibinfo{volume}{95}}, \bibinfo{pages}{177405}
  (\bibinfo{year}{2005}).
\newblock \urlprefix\url{http://dx.doi.org/10.1103/PhysRevLett.95.177405}.

\bibitem{kaer12}
\bibinfo{author}{Kaer, P.}, \bibinfo{author}{Nielsen, T.~R.},
  \bibinfo{author}{Lodahl, P.}, \bibinfo{author}{Jauho, A.-P.} \&
  \bibinfo{author}{M\o{}rk, J.}
\newblock \bibinfo{title}{{Microscopic theory of phonon-induced effects on
  semiconductor quantum dot decay dynamics in cavity QED}}.
\newblock \emph{\bibinfo{journal}{Phys. Rev. B}} \textbf{\bibinfo{volume}{86}},
  \bibinfo{pages}{085302} (\bibinfo{year}{2012}).
\newblock \urlprefix\url{http://dx.doi.org/10.1103/PhysRevB.86.085302}.

\bibitem{madsen13}
\bibinfo{author}{Madsen, K.~H.} \emph{et~al.}
\newblock \bibinfo{title}{{Measuring the effective phonon density of states of
  a quantum dot in cavity quantum electrodynamics}}.
\newblock \emph{\bibinfo{journal}{Phys. Rev. B}} \textbf{\bibinfo{volume}{88}},
  \bibinfo{pages}{045316} (\bibinfo{year}{2013}).
\newblock \urlprefix\url{http://dx.doi.org/10.1103/PhysRevB.88.045316}.

\bibitem{nazir16}
\bibinfo{author}{Nazir, A.} \& \bibinfo{author}{McCutcheon, D. P.~S.}
\newblock \bibinfo{title}{{Modelling exciton-phonon interactions in optically
  driven quantum dots}}.
\newblock \emph{\bibinfo{journal}{J. Phys.: Cond. Mat.}}
  \textbf{\bibinfo{volume}{28}}, \bibinfo{pages}{103002}
  (\bibinfo{year}{2016}).
\newblock \urlprefix\url{http://dx.doi.org/10.1088/0953-8984/28/10/103002}.

\bibitem{borri05}
\bibinfo{author}{Borri, P.} \emph{et~al.}
\newblock \bibinfo{title}{{Exciton dephasing via phonon interactions in InAs
  quantum dots: Dependence on quantum confinement}}.
\newblock \emph{\bibinfo{journal}{Phys. Rev. B}} \textbf{\bibinfo{volume}{71}},
  \bibinfo{pages}{115328} (\bibinfo{year}{2005}).
\newblock \urlprefix\url{http://dx.doi.org/10.1103/PhysRevB.71.115328}.

\bibitem{bayer02}
\bibinfo{author}{Bayer, M.} \& \bibinfo{author}{Forchel, A.}
\newblock \bibinfo{title}{{Temperature dependence of the exciton homogeneous
  linewidth in
  ${\mathrm{In}}_{0.60}{\mathrm{Ga}}_{0.40}\mathrm{As}/\mathrm{GaAs}$
  self-assembled quantum dots}}.
\newblock \emph{\bibinfo{journal}{Phys. Rev. B}} \textbf{\bibinfo{volume}{65}},
  \bibinfo{pages}{041308} (\bibinfo{year}{2002}).
\newblock \urlprefix\url{http://dx.doi.org/10.1103/PhysRevB.65.041308}.

\bibitem{langbein04}
\bibinfo{author}{Langbein, W.} \emph{et~al.}
\newblock \bibinfo{title}{{Radiatively limited dephasing in InAs quantum
  dots}}.
\newblock \emph{\bibinfo{journal}{Phys. Rev. B}} \textbf{\bibinfo{volume}{70}},
  \bibinfo{pages}{033301} (\bibinfo{year}{2004}).
\newblock \urlprefix\url{http://dx.doi.org/10.1103/PhysRevB.70.033301}.

\bibitem{roy16}
\bibinfo{author}{Roy-Choudhury, K.}, \bibinfo{author}{Mann, N.},
  \bibinfo{author}{Manson, R.} \& \bibinfo{author}{Hughes, S.}
\newblock \bibinfo{title}{{Resonance fluorescence spectra from coherently
  driven quantum dots coupled to slow-light photonic crystal waveguides}}.
\newblock \emph{\bibinfo{journal}{Phys. Rev. B}} \textbf{\bibinfo{volume}{93}},
  \bibinfo{pages}{245421} (\bibinfo{year}{2016}).
\newblock \urlprefix\url{http://dx.doi.org/10.1103/PhysRevB.93.245421}.

\bibitem{roy15}
\bibinfo{author}{Roy-Choudhury, K.} \& \bibinfo{author}{Hughes, S.}
\newblock \bibinfo{title}{{Quantum theory of the emission spectrum from quantum
  dots coupled to structured photonic reservoirs and acoustic phonons}}.
\newblock \emph{\bibinfo{journal}{Phys. Rev. B}} \textbf{\bibinfo{volume}{92}},
  \bibinfo{pages}{205406} (\bibinfo{year}{2015}).
\newblock \urlprefix\url{http://dx.doi.org/10.1103/PhysRevB.92.205406}.

\bibitem{kaer13}
\bibinfo{author}{Kaer, P.}, \bibinfo{author}{Lodahl, P.},
  \bibinfo{author}{Jauho, A.-P.} \& \bibinfo{author}{Mork, J.}
\newblock \bibinfo{title}{{Microscopic theory of indistinguishable
  single-photon emission from a quantum dot coupled to a cavity: The role of
  non-Markovian phonon-induced decoherence}}.
\newblock \emph{\bibinfo{journal}{Phys. Rev. B}} \textbf{\bibinfo{volume}{87}},
  \bibinfo{pages}{081308} (\bibinfo{year}{2013}).
\newblock \urlprefix\url{http://dx.doi.org/10.1103/PhysRevB.87.081308}.

\bibitem{rudin06}
\bibinfo{author}{Rudin, S.}, \bibinfo{author}{Reinecke, T.~L.} \&
  \bibinfo{author}{Bayer, M.}
\newblock \bibinfo{title}{{Temperature dependence of optical linewidth in
  single InAs quantum dots}}.
\newblock \emph{\bibinfo{journal}{Phys. Rev. B}} \textbf{\bibinfo{volume}{74}},
  \bibinfo{pages}{161305} (\bibinfo{year}{2006}).
\newblock \urlprefix\url{http://dx.doi.org/10.1103/PhysRevB.74.161305}.

\bibitem{SI_PRL}
\bibinfo{note}{See Supplemental Material at [...] for a detailed description of
  the theory and justification of the used approximations, which includes Refs.
  [59--66].}

\bibitem{takagahara99}
\bibinfo{author}{Takagahara, T.}
\newblock \bibinfo{title}{Theory of exciton dephasing in semiconductor quantum
  dots}.
\newblock \emph{\bibinfo{journal}{Phys. Rev. B}} \textbf{\bibinfo{volume}{60}},
  \bibinfo{pages}{2638} (\bibinfo{year}{1999}).
\newblock \urlprefix\url{http://dx.doi.org/10.1103/PhysRevB.60.2638}.

\bibitem{mahan13}
\bibinfo{author}{Mahan, G.~D.}
\newblock \emph{\bibinfo{title}{{Many-particle physics}}}
  (\bibinfo{publisher}{Springer Science \& Business Media},
  \bibinfo{year}{2013}).

\bibitem{kubo62}
\bibinfo{author}{Kubo, R.}
\newblock \bibinfo{title}{Generalized cumulant expansion method}.
\newblock \emph{\bibinfo{journal}{J. Phys. Soc. Jpn.}}
  \textbf{\bibinfo{volume}{17}}, \bibinfo{pages}{1100} (\bibinfo{year}{1962}).

\bibitem{schwartz16}
\bibinfo{author}{Schwartz, I.} \emph{et~al.}
\newblock \bibinfo{title}{Deterministic generation of a cluster state of
  entangled photons}.
\newblock \emph{\bibinfo{journal}{Science}} \bibinfo{pages}{aah4758}
  (\bibinfo{year}{2016}).
\newblock \urlprefix\url{http://dx.doi.org/10.1126/science.aah4758}.

\bibitem{pryor98}
\bibinfo{author}{Pryor, C.}
\newblock \bibinfo{title}{{Eight-band calculations of strained InAs/GaAs
  quantum dots compared with one-, four-, and six-band approximations}}.
\newblock \emph{\bibinfo{journal}{Phys. Rev. B}} \textbf{\bibinfo{volume}{57}},
  \bibinfo{pages}{7190} (\bibinfo{year}{1998}).
\newblock \urlprefix\url{http://dx.doi.org/10.1103/PhysRevB.57.7190}.

\bibitem{schmidt96}
\bibinfo{author}{Schmidt, K.}, \bibinfo{author}{Medeiros-Ribeiro, G.},
  \bibinfo{author}{Oestreich, M.} \& \bibinfo{author}{Petroff, P.~M.}
\newblock \bibinfo{title}{Excited states in inas self-assembled quantum dots}.
\newblock In \emph{\bibinfo{booktitle}{Photonics West}}, \bibinfo{pages}{185}
  (\bibinfo{organization}{International Society for Optics and Photonics},
  \bibinfo{year}{1996}).
\newblock \urlprefix\url{http://dx.doi.org/10.1117/12.238400}.

\bibitem{noda98}
\bibinfo{author}{Noda, S.}, \bibinfo{author}{Abe, T.} \&
  \bibinfo{author}{Tamura, M.}
\newblock \bibinfo{title}{{Mode assignment of excited states in self-assembled
  InAs/GaAs quantum dots}}.
\newblock \emph{\bibinfo{journal}{Phys. Rev. B}} \textbf{\bibinfo{volume}{58}},
  \bibinfo{pages}{7181} (\bibinfo{year}{1998}).
\newblock \urlprefix\url{http://dx.doi.org/10.1103/PhysRevB.58.7181}.

\bibitem{grundmann95}
\bibinfo{author}{Grundmann, M.}, \bibinfo{author}{Stier, O.} \&
  \bibinfo{author}{Bimberg, D.}
\newblock \bibinfo{title}{{InAs/GaAs pyramidal quantum dots: Strain
  distribution, optical phonons, and electronic structure}}.
\newblock \emph{\bibinfo{journal}{Phys. Rev. B}} \textbf{\bibinfo{volume}{52}},
  \bibinfo{pages}{11969} (\bibinfo{year}{1995}).
\newblock \urlprefix\url{http://dx.doi.org/10.1103/PhysRevB.52.11969}.

\bibitem{landin98}
\bibinfo{author}{Landin, L.}, \bibinfo{author}{Miller, M.},
  \bibinfo{author}{Pistol, M.-E.}, \bibinfo{author}{Pryor, C.} \&
  \bibinfo{author}{Samuelson, L.}
\newblock \bibinfo{title}{{Optical studies of individual InAs quantum dots in
  GaAs: few-particle effects}}.
\newblock \emph{\bibinfo{journal}{Science}} \textbf{\bibinfo{volume}{280}},
  \bibinfo{pages}{262} (\bibinfo{year}{1998}).
\newblock \urlprefix\url{http://dx.doi.org/10.1126/science.280.5361.262}.

\bibitem{stier99}
\bibinfo{author}{Stier, O.}, \bibinfo{author}{Grundmann, M.} \&
  \bibinfo{author}{Bimberg, D.}
\newblock \bibinfo{title}{Electronic and optical properties of strained quantum
  dots modeled by 8-band k\ensuremath{\cdot}p theory}.
\newblock \emph{\bibinfo{journal}{Phys. Rev. B}} \textbf{\bibinfo{volume}{59}},
  \bibinfo{pages}{5688} (\bibinfo{year}{1999}).
\newblock \urlprefix\url{http://dx.doi.org/10.1103/PhysRevB.59.5688}.

\bibitem{santori02}
\bibinfo{author}{Santori, C.}, \bibinfo{author}{Fattal, D.},
  \bibinfo{author}{Vu{\v{c}}kovi{\'c}, J.}, \bibinfo{author}{Solomon, G.~S.} \&
  \bibinfo{author}{Yamamoto, Y.}
\newblock \bibinfo{title}{Indistinguishable photons from a single-photon
  device}.
\newblock \emph{\bibinfo{journal}{Nature}} \textbf{\bibinfo{volume}{419}},
  \bibinfo{pages}{594} (\bibinfo{year}{2002}).
\newblock \urlprefix\url{http://dx.doi.org/10.1038/nature01086}.

\bibitem{kiraz04}
\bibinfo{author}{Kiraz, A.}, \bibinfo{author}{Atat\"ure, M.} \&
  \bibinfo{author}{Imamo\ifmmode~\breve{g}\else \u{g}\fi{}lu, A.}
\newblock \bibinfo{title}{{Quantum-dot single-photon sources: Prospects for
  applications in linear optics quantum-information processing}}.
\newblock \emph{\bibinfo{journal}{Phys. Rev. A}} \textbf{\bibinfo{volume}{69}},
  \bibinfo{pages}{032305} (\bibinfo{year}{2004}).
\newblock \urlprefix\url{http://dx.doi.org/10.1103/PhysRevA.69.032305}.

\bibitem{johansen08}
\bibinfo{author}{Johansen, J.} \emph{et~al.}
\newblock \bibinfo{title}{Size dependence of the wavefunction of self-assembled
  inas quantum dots from time-resolved optical measurements}.
\newblock \emph{\bibinfo{journal}{Phys. Rev. B}} \textbf{\bibinfo{volume}{77}},
  \bibinfo{pages}{073303} (\bibinfo{year}{2008}).
\newblock \urlprefix\url{http://dx.doi.org/10.1103/PhysRevB.77.073303}.

\bibitem{madsen14}
\bibinfo{author}{Madsen, K.~H.} \emph{et~al.}
\newblock \bibinfo{title}{Efficient out-coupling of high-purity single photons
  from a coherent quantum dot in a photonic-crystal cavity}.
\newblock \emph{\bibinfo{journal}{Phys. Rev. B}} \textbf{\bibinfo{volume}{90}},
  \bibinfo{pages}{155303} (\bibinfo{year}{2014}).
\newblock \urlprefix\url{http://dx.doi.org/10.1103/PhysRevB.90.155303}.

\bibitem{bennett16}
\bibinfo{author}{Bennett, A.~J.} \emph{et~al.}
\newblock \bibinfo{title}{Cavity-enhanced coherent light scattering from a
  quantum dot}.
\newblock \emph{\bibinfo{journal}{Science Advances}}
  \textbf{\bibinfo{volume}{2}}, \bibinfo{pages}{e1501256}
  (\bibinfo{year}{2016}).
\newblock \urlprefix\url{http://dx.doi.org/10.1126/sciadv.1501256}.

\bibitem{werschler16}
\bibinfo{author}{Werschler, F.} \emph{et~al.}
\newblock \bibinfo{title}{{Coupling of Excitons and Discrete Acoustic Phonons
  in Vibrationally Isolated Quantum Emitters}}.
\newblock \emph{\bibinfo{journal}{Nano Lett.}} \textbf{\bibinfo{volume}{16}},
  \bibinfo{pages}{5861} (\bibinfo{year}{2016}).
\newblock \urlprefix\url{http://dx.doi.org/10.1021/acs.nanolett.6b02667}.

\bibitem{masumoto13}
\bibinfo{author}{Masumoto, Y.} \& \bibinfo{author}{Takagahara, T.}
\newblock \emph{\bibinfo{title}{{Semiconductor quantum dots: physics,
  spectroscopy and applications}}} (\bibinfo{publisher}{Springer Science \&
  Business Media}, \bibinfo{year}{2013}).

\bibitem{coles16}
\bibinfo{author}{Coles, R.} \emph{et~al.}
\newblock \bibinfo{title}{Chirality of nanophotonic waveguide with embedded
  quantum emitter for unidirectional spin transfer}.
\newblock \emph{\bibinfo{journal}{Nat. Commun.}} \textbf{\bibinfo{volume}{7}}
  (\bibinfo{year}{2016}).
\newblock \urlprefix\url{http://dx.doi.org/10.1038/ncomms11183}.

\bibitem{heiss13}
\bibinfo{author}{Heiss, M.} \emph{et~al.}
\newblock \bibinfo{title}{Self-assembled quantum dots in a nanowire system for
  quantum photonics}.
\newblock \emph{\bibinfo{journal}{Nat. Mater.}} \textbf{\bibinfo{volume}{12}},
  \bibinfo{pages}{439} (\bibinfo{year}{2013}).
\newblock \urlprefix\url{http://dx.doi.org/10.1038/nmat3557}.

\bibitem{anghel07}
\bibinfo{author}{Anghel, D.~V.} \& \bibinfo{author}{K\"{u}hn, T.}
\newblock \bibinfo{title}{Quantization of the elastic modes in an isotropic
  plate}.
\newblock \emph{\bibinfo{journal}{J. Phys. A: Mathematical and Theoretical}}
  \textbf{\bibinfo{volume}{40}}, \bibinfo{pages}{10429} (\bibinfo{year}{2007}).
\newblock \urlprefix\url{http://dx.doi.org/10.1088/1751-8113/40/34/003}.

\bibitem{midolo12}
\bibinfo{author}{Midolo, L.} \emph{et~al.}
\newblock \bibinfo{title}{Electromechanical tuning of vertically-coupled
  photonic crystal nanobeams}.
\newblock \emph{\bibinfo{journal}{Opt. Express}} \textbf{\bibinfo{volume}{20}},
  \bibinfo{pages}{19255} (\bibinfo{year}{2012}).
\newblock
  \urlprefix\url{http://www.opticsexpress.org/abstract.cfm?URI=oe-20-17-19255}.

\bibitem{reigue16}
\bibinfo{author}{Reigue, A.} \emph{et~al.}
\newblock \bibinfo{title}{Probing electron-phonon interaction through
  two-photon interference in resonantly driven quantum dots}.
\newblock \emph{\bibinfo{journal}{Phys. Rev. Lett.}}
  \textbf{\bibinfo{volume}{118}}, \bibinfo{pages}{233602}
  (\bibinfo{year}{2017}).
\newblock \urlprefix\url{https://dx.doi.org/10.1103/PhysRevLett.118.233602}.

\bibitem{fry00}
\bibinfo{author}{Fry, P.~W.} \emph{et~al.}
\newblock \bibinfo{title}{{Inverted Electron-Hole Alignment in InAs-GaAs
  Self-Assembled Quantum Dots}}.
\newblock \emph{\bibinfo{journal}{Phys. Rev. Lett.}}
  \textbf{\bibinfo{volume}{84}}, \bibinfo{pages}{733} (\bibinfo{year}{2000}).

\bibitem{skinner86}
\bibinfo{author}{Skinner, J.~L.} \& \bibinfo{author}{Hsu, D.}
\newblock \bibinfo{title}{{Pure dephasing of a two-level system}}.
\newblock \emph{\bibinfo{journal}{J. Phys. Chem.}}
  \textbf{\bibinfo{volume}{90}} (\bibinfo{year}{1986}).
\newblock \urlprefix\url{http://dx.doi.org/10.1021/j100412a013}.

\bibitem{tighineanu15}
\bibinfo{author}{Tighineanu, P.}, \bibinfo{author}{S\o{}rensen, A.~S.},
  \bibinfo{author}{Stobbe, S.} \& \bibinfo{author}{Lodahl, P.}
\newblock \bibinfo{title}{Unraveling the mesoscopic character of quantum dots
  in nanophotonics}.
\newblock \emph{\bibinfo{journal}{Phys. Rev. Lett.}}
  \textbf{\bibinfo{volume}{114}}, \bibinfo{pages}{247401}
  (\bibinfo{year}{2015}).
\newblock
  \urlprefix\url{https://link.aps.org/doi/10.1103/PhysRevLett.114.247401}.

\bibitem{tighineanu13}
\bibinfo{author}{Tighineanu, P.} \emph{et~al.}
\newblock \bibinfo{title}{Decay dynamics and exciton localization in large
  {GaAs} quantum dots grown by droplet epitaxy}.
\newblock \emph{\bibinfo{journal}{Phys. Rev. B}} \textbf{\bibinfo{volume}{88}},
  \bibinfo{pages}{155320} (\bibinfo{year}{2013}).
\newblock \urlprefix\url{https://link.aps.org/doi/10.1103/PhysRevB.88.155320}.

\bibitem{vurgaftman01}
\bibinfo{author}{Vurgaftman, I.}, \bibinfo{author}{Meyer, J.~R.} \&
  \bibinfo{author}{Ram-Mohan, L.~R.}
\newblock \bibinfo{title}{Band parameters for iii--v compound semiconductors
  and their alloys}.
\newblock \emph{\bibinfo{journal}{Journal of Applied Physics}}
  \textbf{\bibinfo{volume}{89}}, \bibinfo{pages}{5815--5875}
  (\bibinfo{year}{2001}).
\newblock \urlprefix\url{https://doi.org/10.1063/1.1368156}.
\newblock \eprint{https://doi.org/10.1063/1.1368156}.

\bibitem{auld73}
\bibinfo{author}{Auld, B.~A.}
\newblock \emph{\bibinfo{title}{Acoustic fields and waves in solids}}
  (\bibinfo{publisher}{John Wiley \& Sons}, \bibinfo{year}{1973}).

\bibitem{stroscio94}
\bibinfo{author}{Stroscio, M.~A.}, \bibinfo{author}{Kim, K.~W.},
  \bibinfo{author}{Yu, S.} \& \bibinfo{author}{Ballato, A.}
\newblock \bibinfo{title}{Quantized acoustic phonon modes in quantum wires and
  quantum dots}.
\newblock \emph{\bibinfo{journal}{J. Appl. Phys.}}
  \textbf{\bibinfo{volume}{76}}, \bibinfo{pages}{4670} (\bibinfo{year}{1994}).

\bibitem{LL70}
\bibinfo{author}{Landau, L.} \& \bibinfo{author}{Lifshitz, E.}
\newblock \emph{\bibinfo{title}{Theory of Elasticity (2nd ed.)}}
  (\bibinfo{publisher}{Pergamon Press}, \bibinfo{year}{1970}).

\end{thebibliography}

\newpage

\setcounter{figure}{0}
\renewcommand{\thefigure}{S\arabic{figure}}
\setcounter{equation}{0}
\renewcommand{\theequation}{S\arabic{equation}}

\clearpage
\onecolumngrid

\section{Theory of linear and quadratic coupling to acoustic phonons in nano-structures}
\label{sec:Theory_Coupling_to_Acoustic_Phonons}
In this work we generalize the formalism presented in Ref.~\cite{muljarov04} to nano-structures and perform a key approximation, the Gaussian approximation to the phonon bath, which we show below to be well justified for QDs and results in simple results. We assume time-reversal symmetry, which implies that all the quantities (wave functions, phonon modes, etc.) can be chosen to be real. This means that the theory can be employed for single-particle excitations like excitons, trions, etc., as long as no fields or processes that destroy the time-reversal symmetry (e.g., magnetic fields) are present. We consider phonon processes that happen within the same spin manifold. This means that we neglect weak spin-flip processes between, e.g., bright-bright and bright-dark excitons.

We consider an exciton in a QD coupled to acoustic phonons with the following Hamiltonian~\cite{muljarov04}
\begin{equation}
\begin{split}
H &= \sum_nE_n\proj{\psi_n}{\psi_n} + \sum_\mathbf{q}\hbar\omega_\mathbf{q}\ad_\mathbf{q}a_\mathbf{q} + \sum_{nm}V_{nm}\proj{\psi_n}{\psi_m},\\
V_{nm} &= \sum_\mathbf{q}L_\mathbf{q}^{nm}(a_\mathbf{q}+\ad_\mathbf{q}),
\end{split}
\label{eq:Hamiltonian_general}
\end{equation}
where the first two terms correspond to the bare Hamiltonians of the QD and the phonons, respectively, and the third term is the interaction Hamiltonian. Here, $E_n$ is the energy of the $n$-th QD eigenstate $\psi_n$, $\ad_\qq$ ($a_\qq$) is the creation (annihilation) operator of the phonon mode with wavevector $\qq$ and energy $\hbar\omega_\qq$, and $L_\qq^{nm}$ is the effective electron-phonon matrix element. In writing the above equation, we have chosen the phase convention such that $L_\mathbf{q}^{nm}$ is real. The interaction with optical phonons is neglected since it does not play a role at low temperatures.

Two mechanisms dominate the interaction between electrons and acoustic phonons: deformation-potential and piezo-electric coupling~\cite{takagahara99}. The former describes the deformation of the lattice induced by the creation of an electron-hole pair inside the QD and has a matrix element of the form~\cite{takagahara99}
\begin{equation}
L_{\mathbf{q},\mathrm{DF}}^{mn} = D_e\bracket{\psi_m}{\divv \mathbf{u}_\mathbf{q}(\rr_e)}{\psi_n} - D_h\bracket{\psi_m}{\divv \mathbf{u}_\mathbf{q}(\rr_h)}{\psi_n},
\label{eq:def_pot}
\end{equation}
where $\mathbf{u}_\qq$ is the phonon-displacement mode, and $D_e$ ($D_h$) is the deformation-potential constant of electrons (holes). Equation~(\ref{eq:def_pot}) describes the mechanical deformation of the crystal lattice induced by the creation of an electron-hole pair inside the QD. On the other hand, the piezo-electric coupling describes the interaction between the exciton and the polarization of the crystal lattice~\cite{takagahara99}
\begin{equation}
L_{\mathbf{q},\mathrm{PZ}}^{mn} = \frac{2e p}{\epsilon_0\epsilon}\hat{\mathbf{G}}\cdot\left(\bracket{\psi_m}{\mathbf{u}_\mathbf{q}(\rr_e)}{\psi_n} - \bracket{\psi_m}{\mathbf{u}_\mathbf{q}(\rr_h)}{\psi_n}\right),
\label{eq:piezo-electric}
\end{equation}
where $\epsilon_0\epsilon$ ($p$) is the dielectric (piezo-electric) constant and $\hat{\mathbf{G}}$ is a geometric operator that is of the order of unity.

In the following we argue that the piezo-electric interaction is negligible in our study. Two types of phonons dephase the interaction between QDs and light in nano-structures: short-wavelength phonons of the order of the QD size $\lambda_\mathrm{ph}\sim L$, and long-wavelength phonons $\lambda\rightarrow\infty$ that arise in 1D and 2D structures. It is shown in Ref.~\cite{takagahara99} that the short-wavelength phonons dephase the QD mainly through the deformation-potential coupling with a negligible contribution from the piezo-electric coupling. To show that the piezo-electric coupling is negligible also for long-wavelength deformations, we expand the mode displacement, $u_\qq \propto \exp(\I \qq\cdot\rr)$, in a Taylor series with respect to the electron (hole) center of mass, $\rr_e$ ($\rr_h$). Using Eqs.~(\ref{eq:def_pot}) and (\ref{eq:piezo-electric}) this yields
\begin{equation}
\abs{\frac{L_{\mathbf{q},\mathrm{PZ}}^{11}}{L_{\mathbf{q},\mathrm{DF}}^{11}}}^2 \simeq \abs{\frac{2ep}{\epsilon_0\epsilon}\frac{\bracket{\psi_1}{\mathbf{r}_e-\mathbf{r}_h}{\psi_1}}{D_e-D_h}}^2 \approx 10^{-2},
\end{equation}
where typical GaAs values were used $D_e=-\SI{14.6}{\electronvolt}$, $D_h=-\SI{4.8}{\electronvolt}$, $p=\SI{0.16}{\coulomb\meter^2}$, $\epsilon = 12.56$~\cite{takagahara99}, and a typical electron-hole separation of $\sim \SI{0.4}{\nano\meter}$ in In(Ga)As QDs~\cite{fry00}. We have only considered the contribution to $L_\qq^{11}$ because the other matrix elements are not relevant at small phonon energies as is shown later. We therefore neglect the piezo-electric coupling and consider $L_{\mathbf{q}}^{mn}\equiv L_{\mathbf{q},\mathrm{DF}}^{mn}$ in the following.

We assume that at $t=0$ an exciton is created in the ground state $\ket{\psi_1}$, and seek to calculate the evolution of the QD coherence $P(t)=i\left<\sigma^-(t)\sigma^+(0)\right>$ at $t>0$. Solving for $P(t)$ in the most general case with the Hamiltonian in \eqref{eq:Hamiltonian_general} is highly challenging. The problem can be simplified by noting that, at low temperature, only the ground state $\ket{\psi_1}$ of the QD is populated. The excited states can then be eliminated as explained in Ref.~\cite{muljarov04}, so that only virtual transition to excited states are taken into account. The resulting simplified Hamiltonian reads
\begin{equation}
\begin{split}
H &= H_0 + (V_\mathrm{L}+V_\mathrm{Q})\ket{\psi_1}\bra{\psi_1},\\
V_\mathrm{L} &= \sum_\qq L_\qq A_\qq,\hspace{0.3cm} V_\mathrm{Q} = \sum_{b,m}\left[\sum_{\qq}Q_{\qq b}^{m}A_\qq\right]^2,\\
L_\qq \equiv L_\qq^{11} &= M_{\qq e}^{11} - M_{\qq h}^{11}, \hspace{1cm} Q_{\qq b}^{m} = \frac{M_{\qq b}^{1m}}{\sqrt{\Delta_m}},
\end{split}
\label{eq:V_L_V_Q}
\end{equation}
where $A_\qq=a_\qq+\ad_\qq$, $b = \{\mathrm{e},\mathrm{h}\}$ denotes an electron or hole, $M_{\qq b}^{mn}=D_b \me{\psi_{m b}}{\divv \mathbf{u}_\qq}{\psi_{n b}}$, and $\Delta_m$ is the energy distance between the ground, $\ket{\psi_{1 b}}$, and $m$-th excited state of the QD, $\ket{\psi_{m b}}$. The Hamiltonian is now diagonal in the QD subspace but is quadratic in phonon displacement. Solving for $P(t)$ can be done using the cumulant expansion method presented in Refs.~\cite{skinner86,muljarov04} with the key difference that the phonon factorization used in those works does not hold in the case of a nano-structure. In the following we explain how to deal with this complication.

We start by writing $P(t)$ as~\cite{mahan13}
\begin{equation}
P(t) = \left< \mathcal{T}\mathrm{e}^{ -\frac{\I}{\hbar}\int_0^t\mathrm{d}\tau \left(V_\mathrm{L}+V_\mathrm{Q}\right) } \right>,
\end{equation}
where $\mathcal{T}$ denotes the time-ordering operator. Denoting $F(t)=-(i/\hbar)\int_0^t\mathrm{d}\tau \left(V_\mathrm{L}+V_\mathrm{Q}\right)$ and employing a well-known theorem for the cumulant~\cite{kubo62} leads to
\begin{equation}
\begin{split}
P(t) &= \mathrm{exp}\left\{ \sum_{n=1}^\infty \frac{1}{n!}\av{\mathcal{T}F^n(t)}_\mathrm{conn} \right\} \\
&= \mathrm{exp}\left\{\sum_{n=1}^\infty \left(\frac{-\I}{\hbar}\right)^n \frac{1}{n!} \int_0^t\mathrm{d}t_1\int_0^t\mathrm{d}t_2\ldots\int_0^t\mathrm{d}t_n\av{\mathcal{T}V(t_1)V(t_2)\ldots V(t_n)}_\mathrm{conn}\right\},
\end{split}
\label{eq:P_t_general}
\end{equation}
where $V=V_\mathrm{L}+V_\mathrm{Q}$, and the averaging is now performed over connected diagrams only. 
Let us assume for the moment that the linear, $V_\mathrm{L}$, and quadratic, $V_\mathrm{Q}$, potentials do not mix (we return to the $V_\mathrm{L}$--$V_\mathrm{Q}$ interference terms at the end of this section and argue that they do not play a role) allowing to express $P(t) = \exp\left[K_\mathrm{L}(t)+K_\mathrm{Q}(t)\right]$ with
\begin{align}
K_\mathrm{L}(t) &= \sum_{n=1}^\infty \left(\frac{-\I}{\hbar}\right)^n \frac{1}{n!} \int_0^t\mathrm{d}t_1\int_0^t\mathrm{d}t_2\ldots\int_0^t\mathrm{d}t_n\av{\mathcal{T}V_\mathrm{L}(t_1)V_\mathrm{L}(t_2)\ldots V_\mathrm{L}(t_n)}_\mathrm{conn},\\
K_\mathrm{Q}(t) &= \sum_{n=1}^\infty \left(\frac{-\I}{\hbar}\right)^n \frac{1}{n!} \int_0^t\mathrm{d}t_1\int_0^t\mathrm{d}t_2\ldots\int_0^t\mathrm{d}t_n\av{\mathcal{T}V_\mathrm{Q}(t_1)V_\mathrm{Q}(t_2)\ldots V_\mathrm{Q}(t_n)}_\mathrm{conn}.
\end{align}
It can be easily checked that the linear interaction contains only one connected diagram, $n=2$, yielding
\begin{align}
K_\mathrm{L}(t) &= -\frac{\I}{2\hbar}\sum_{\qq}\abs{L_\qq}^2\iint_0^t\mathrm{d}t_1\mathrm{d}t_2 D_\qq(t_1-t_2),\\
D_\qq(t) &= -\frac{\I}{\hbar}\av{\mathcal{T}A_\qq(t_1)A_\qq(t_2)}_\mathrm{conn} = -\frac{\I}{\hbar}\left[ (N_\qq+1)\mathrm{e}^{-\I\omega_\qq\abs{t}} + N_\qq\mathrm{e}^{\I\omega_\qq\abs{t}} \right],
\end{align}
where $D_\qq(t)$ is the phonon Green function~\cite{mahan13}. Performing the time integration leads to
\begin{align}
\label{eq:KL_t}
K_\mathrm{L}(t) &= -\frac{1}{2\hbar^2}\sum_\qq \abs{L_\qq}^2d_\qq(t),\\
d_\qq(t)&=(2N_\qq+1) \left(\frac{\sin\frac{\omega_\qq t}{2}}{\frac{\omega_\qq}{2}}\right)^2 + \frac{2\I}{\omega_\qq^2}\left( \sin\omega_\qq t - \omega_\mathbf{q}t \right).
\end{align}
This is the main result of the independent boson model~\cite{mahan13} and is well known in the context of a bulk medium, where it leads to broad sidebands in the spectrum $S(\omega) = \mathrm{Re}\int_0^\infty\mathrm{d}tP(t)\exp(-\I\omega t)$. The linear cumulant contains a single propagator corresponding to the emission or absorption of a phonon, see Fig.~\ref{fig:exact_vs_perturbation}(a). 

The derivation of the quadratic coupling, $K_\mathrm{Q}(t)$, is more complicated and relies on a diagrammatic representation of the cumulant~\cite{skinner86}, which is outlined in the following. We drop the electron/hole index, $b$, to simplify the notation. We first note that the contribution from $n=1$ to $K_\mathrm{Q}(t)$ results in a quantity that is linearly varying in time, and is denoted as $-\I\mu_F$ in the main text. Taking $n=2$ gives two identical connected diagrams and yields
\begin{equation}
\begin{split}
K_\mathrm{Q}^{n=2}(t) &= \sum_{m_1m_2}\sum_{\qq_1}Q_{\qq_1}^{m_1}Q_{\qq_1}^{m_2}\sum_{\qq_2}Q_{\qq_2}^{m_1}Q_{\qq_2}^{m_2}\iint \mathrm{d}t_1\mathrm{d}t_2 D_{\qq_1}(t_1-t_2)D_{\qq_2}(t_2-t_1)\\
&= \sum_{m_1m_2}\iint\mathrm{d}t_1\mathrm{d}t_2\left[ Q_\qq^{m_1}Q_\qq^{m_2}D_\qq(t_1-t_2) \right]^2,
\end{split}
\label{eq:KQ_n_2}
\end{equation}
which is equivalent to the third line of Eq.~(5) of the main text. Generalizing this approach for $n=N$ can be done by identifying the number of equivalent connected diagrams~\cite{skinner86,muljarov04}. We thus obtain
\begin{align}
\label{eq:KQ_general}
K_\mathrm{Q}(t) &= \frac{1}{2}\sum_{n=1}^\infty \sum_{m_1\ldots m_n}\frac{1}{n} \iiint\ldots\int\mathrm{d}t_1\mathrm{d}t_2\ldots\mathrm{d}t_n D_\mathrm{Q}^{m_1m_2}(t_1-t_2)D_\mathrm{Q}^{m_2m_3}(t_2-t_3)\ldots D_\mathrm{Q}^{m_nm_1}(t_n-t_1),\\
D_\mathrm{Q}^{mn}(t) &= 2 \sum_\qq Q_\qq^{m}Q_\qq^{n}D_\qq(t).
\end{align}
Equation (\ref{eq:KQ_general}) is a generalization of the result from Ref.~\cite{muljarov04} for the case when no factorization with respect to the QD states can be performed. The quadratic cumulant contains an infinite sum of connected diagrams as illustrated in Fig.~\ref{fig:exact_vs_perturbation}(b), and can be evaluated numerically at each time $t$ using the Fredholm eigenvalue problem that is presented in Ref.~\cite{muljarov04}. However, we find that the interaction between QDs and phonons is sufficiently weak such that most of the physics is contained in the term $n=2$ from \eqref{eq:KQ_n_2} as shown in the next section.

We now discuss the terms arising from the mixing of $V_\mathrm{L}$ and $V_\mathrm{Q}$ in the connected diagram of \eqref{eq:P_t_general}, and argue that they do not play a role in this study. Since the contribution to \eqref{eq:P_t_general} is only from connected diagrams, the only non-vanishing mixing terms are the processes in which an arbitrary number of scattering events stemming from the quadratic coupling are sandwiched between an emission/absorption event that stems from the linear interaction. All the other diagrams vanish identically either because they contain an odd number of creation/annihilation phonon operators or because they do not form connected diagrams. This yields the following mixed cumulant $K_\mathrm{M}(t)$
\begin{equation}
K_\mathrm{M}(t) = \sum_{n=3}^\infty \left(\frac{-\I}{\hbar}\right)^n \frac{1}{n!} \int_0^t\mathrm{d}t_1\int_0^t\mathrm{d}t_2\ldots\int_0^t\mathrm{d}t_n\av{\mathcal{T}V_\mathrm{L}(t_1)V_\mathrm{Q}(t_2)\ldots V_\mathrm{Q}(t_{n-1})V_\mathrm{L}(t_n)}_\mathrm{conn}.
\end{equation}
This process corresponds to a phonon that is created, scatters $t_{n-2}$ times, and is then absorbed by the QD.
We consider spherical QDs in this study meaning that the wavefunctions are parity symmetric (i.e., even or odd). It thus follows that $K_\mathrm{M}(t)$ vanishes in parity-symmetric environments because $V_\mathrm{L}$ and $V_\mathrm{Q}$ contain matrix elements that are orthogonal, i.e., $V_\mathrm{L}$ couples to modes of even parity whereas $V_\mathrm{Q}$ couples to modes that are odd. 
Most of the structures studied in this work are parity symmetric and $K_\mathrm{M}(t)$ is not relevant. The only exception is the study of decoherence in photonic waveguides with QDs positioned away from the cross-sectional center in Sec.~\ref{sec:1D_structures}. In this case, however, the contribution from the linear interaction in \eqref{eq:KL_t} is a lower order process and is dominant at low temperature. We thus neglect $K_\mathrm{M}(t)$ in this work.

More generally, the mixed cumulant $K_\mathrm{M}(t)$ is non-negligible if (i) the QDs are positioned within a few nanometers from the interface where phonons of the size of the QD are affected by the interface and are not parity symmetric, and (ii) the QD wavefunctions are not parity symmetric. The former situation does not occur in realistic photonic devices since QDs are normally placed many tens of nanometers away from surfaces for optimal performance. However, the latter condition may occur in practice. Recent research suggests that In(Ga)As QDs may possess wavefunctions that lack parity symmetry~\cite{tighineanu15}. Even if such a mixing term exists, it is of higher order than the coupling considered here. We thus neglect this complication in this study and consider spherical QDs as explained in detail in the following section.

\section{Modeling the quantum dot}
\label{sec:modeling_QD}
Microscopically, the interaction between the QD and phonons happens through a matrix element involving the exciton wave functions and the curl-free component of the phonon displacement. The size and shape of the wave functions therefore play an important role in the strength of this interaction and thus in the resulting dephasing. An exception is the linear-coupling-broadening of the zero-phonon line in nano-structures, which is independent of the QD size. Here we assume the strong-confinement regime in which the electrons and holes are spatially uncorrelated and can be modeled separately, which is motivated by the small size of typical In(Ga)As QDs compared to the exciton Bohr radius. We consider a spherical QD with parabolic confinement and the following wave functions that are the same for electrons and holes
\begin{equation}
\begin{split}
\psi_{1b} &= \frac{1}{\pi^{3/4}L^{3/2}}\mathrm{e}^{-r^2/2L^2},\\
\psi_{2b} &= \frac{2\sqrt{2}}{\sqrt{3}\pi^{1/4}L^{5/2}}Y_1^m(\theta,\phi)r\mathrm{e}^{-r^2/2L^2},
\end{split}
\end{equation}
where $r$ is the radial coordinate, $L$ the radius of the QD wave function, and $Y_l^m(\theta,\phi)$ the spherical harmonic. There are three excited states degenerate in energy with $m=\{-1, 0, 1\}$. The linear exciton-phonon coupling is solely mediated by the ground-state wave function, while the quadratic coupling is determined by the phonon-mediated interaction between the ground and excited wave functions.

\section{Spectral filtering and source efficiency}
\label{sec:spectral_filtering_and_source_efficiency}
The emission spectrum of QDs (Fig.~1(a) in the main text) consists of broad spectral sidebands, which originate from the emission or absorption of a phonon over pico-second time scales, and a narrow zero-phonon line, which is broadened by phonon processes occurring over nano-second time scales. The former is fully incoherent while the latter is partly coherent with a proportion that depends on temperature and the parameters of the nano-structure. To enhance coherence, the phonon sidebands are commonly removed in experiments at the expense of the emitter efficiency. In this work we thus only study the part of the phonon decoherence that affects the experimentally relevant zero-phonon line. In the following we discuss the practicability and impact of the spectral filtering on the emitter efficiency.

We first note that the filtering is relatively straightforward to conduct~\cite{somaschi16} due to the large spectral mismatch between the sidebands and the zero-phonon line. The spectral width of the phonon sidebands is of the order of $v_s 2\pi / L \approx 2\pi\times\SI{1}{\tera\hertz}$ for realistic QD sizes, where $v_s$ is the longitudinal speed of sound, while the zero-phonon line is of the order of the QD natural linewidth, $\sim 2\pi\times\SI{160}{\mega\hertz}$, but can vary slightly depending on the Purcell enhancement of light-matter interaction or the amount of phonon decoherence. There may be further sources of decoherence such as charge noise but these can be efficiently neutralized experimentally even in nano-structures~\cite{kuhlmann15, kirsanske17} and we thus do not consider them here. A spectral filter of, e.g., $2\pi\times\SI{5}{\giga\hertz}$ would therefore remove the sidebands almost entirely while letting the zero-phonon line through. The source efficiency is decreased to about $\sim85-\SI{95}{\percent}$ depending on whether the QD is embedded in a cavity as argued in Ref.~\cite{IlesSmith17}. This calculation was conducted for bulk phonons but we expect negligible difference for the nano-structures we consider because the phonon sidebands are negligibly affected, see the discussion in the first paragraph of Sec.~\ref{sec:3D_and_single_scattering_approx}.

\section{Choice of parameters}
\label{sec:choice_of_parameters}
Our theory of phonon decoherence models microscopically the interaction between a QD and the surrounding phonon bath. The resulting expressions for the decoherence contain a number of physical parameters describing properties of the QD and the phonons. Some of these parameters are empirically well established, while others have larger uncertainties. In the following we justify the choice of the parameters used in this paper.

The parameter with the largest uncertainty in our theory is the radius of the QD wave function, $L$. To our knowledge, no experiment has reported a direct measurement of the QD wave function. Calculating this value theoretically is non-trivial either due to the largely unknown microscopic profile of the QDs. We therefore have to rely on indirect estimates of the size. In Ref.~\cite{muljarov04} such an estimate yields a radius the an In(Ga)As QD wave function of $\sim\SI{3}{\nano\meter}$ by fitting to experimental data. In Ref.~\cite{madsen13} a similar estimate yields values between $\sim\SI{1.5}{\nano\meter}$ and $\sim\SI{4}{\nano\meter}$ depending on the confinement direction. In a different material system (GaAs QDs) with similar electron and hole effective masses, values between \SI{1.9}{\nano\meter} and \SI{3.6}{\nano\meter} were found~\cite{tighineanu13}. In addition to the uncertainties in extracting the parameters, the values will likely change from QD to QD due to the random character of the self-assembled growth procedure. To take all these considerations into account, we perform the calculations for several different QD radii in Figs. 1 and 2 that will likely encompass most experimental values: \SI{1.5}{\nano\meter}, \SI{3}{\nano\meter}, and \SI{4.5}{\nano\meter}. In Figs. 3 and 4, we use a QD radius of \SI{3}{\nano\meter}. We note that we complement these results with analytic expressions that could be readily evaluated for the desired microscopic parameters.

The deformation potentials for the conduction and valence bands are another source of uncertainty -- values that differ by at least a factor of two are reported in the literature~\cite{vurgaftman01}. This only affects the evaluation of the quadratic exciton-phonon coupling. Since the corresponding dephasing rate scales drastically as $T^{11}$ at low temperatures and only as the fourth power of the deformation potentials, these parameter uncertainties will not interfere with the main message of our paper. The linear exciton-phonon coupling describes how the band gap changes under stress and therefore depends on the difference of the deformation potentials. Since changes in the band gap are easier to measure, this difference is known with a higher accuracy. Here we use the values employed in Ref.~\cite{lindwall07}.

Another parameter that is poorly known is the energy distance between the states $\ket{1}$ and $\ket{2}$ for electrons and holes in the QD. Due to the different effective masses, the electrons are quantized stronger than the holes. Here we assume a quantization energy for electrons twice as large $\Delta_e=2\Delta_h$, which qualitatively agrees with previous calculations and experimental values~\cite{schwartz16, pryor98, schmidt96, noda98, grundmann95, landin98, stier99}. Following the same references, and noting that for an infinite potential well the quantization energy scales as $L^{-n}$ with $n=2$ while for a shallow well $n$ is only slightly above one, we assume a dependence of the form $\Delta_e = \SI{40}{\milli\electronvolt}\times \SI{3}{\nano\meter}/L$. Importantly, this energy difference is only relevant for the evaluation of the quadratic exciton-phonon coupling. Since the dephasing rate scales as $\Delta^{-2}$ compared to $T^{11}$, uncertainties related to this parameter will not affect the main message of the paper.

The rest of the parameters such as mass density and speed of sound are well established. We provide the corresponding references in the main text.

\section{Bulk media (3D) and the Gaussian approximation}
\label{sec:3D_and_single_scattering_approx}
In the following we evaluate $P(t)=\exp(K_\mathrm{L}(t)+K_\mathrm{Q}(t))$ in a bulk medium. In this case the phonon modes are propagating plane waves, $\divv\textbf{u}_\qq(\rr)=\sqrt{\hbar\omega_\qq/2\rho_m v_s^2  V}\exp(\mathrm{i}\qq\cdot\rr)$, where $\rho_m$ is the mass density, $v_s$ the longitudinal speed of sound, and $V$ the quantization volume. 
We consider a spherical QD with parabolic confinement potential as explained in the main text. Plugging $\divv\textbf{u}_\qq$ into \eqref{eq:KL_t} yields
\begin{equation}
K_\mathrm{L}^{\mathrm{3D}}(t) = -\frac{(D_e-D_h)^2}{8\pi^2\hbar\rho_mv_s^3L^2}\left\{\int_0^\infty 2(1+2N_q)\left[1-\cos(qLt/t_0)\right](qL)\mathrm{e}^{-(qL)^2/2}\mathrm{d}(qL) - \mathrm{i}\sqrt{2\pi}\frac{t}{t_0}\left( 1 - \mathrm{e}^{-(t/t_0)^2/2} \right)\right\},
\end{equation}
and the corresponding linear coherence function is plotted in Fig.~2(a) in the main text. The corresponding spectrum consists of broad phonon sidebands depicted in Fig.~1(a) in the main text. In a modified phononic environment, the phonon sidebands mimic the joint phonon density of states of the particular dimensionality of the structure. For instance, the sidebands are a series of sharp lines in a 0D geometry, see Fig.~\ref{fig:sphere_P_ZPL}. Similarly, in a photonic waveguide the sidebands acquire 1D-like satellite peaks as shown in Ref.~\cite{lindwall07}. However, since QDs interact only with modes with a wavelength larger than the QD size, the phonon sidebands are prominent only in small photonic structures with a discrete number of such modes. In structures larger than 40--50 nm, the relevant phonon wavelengths are almost unaffected by the environment so that many modes are interacting with the QDs and the sidebands quickly approach the bulk limit.

\begin{figure}[t!]
\centering
\includegraphics[width=0.8\columnwidth]{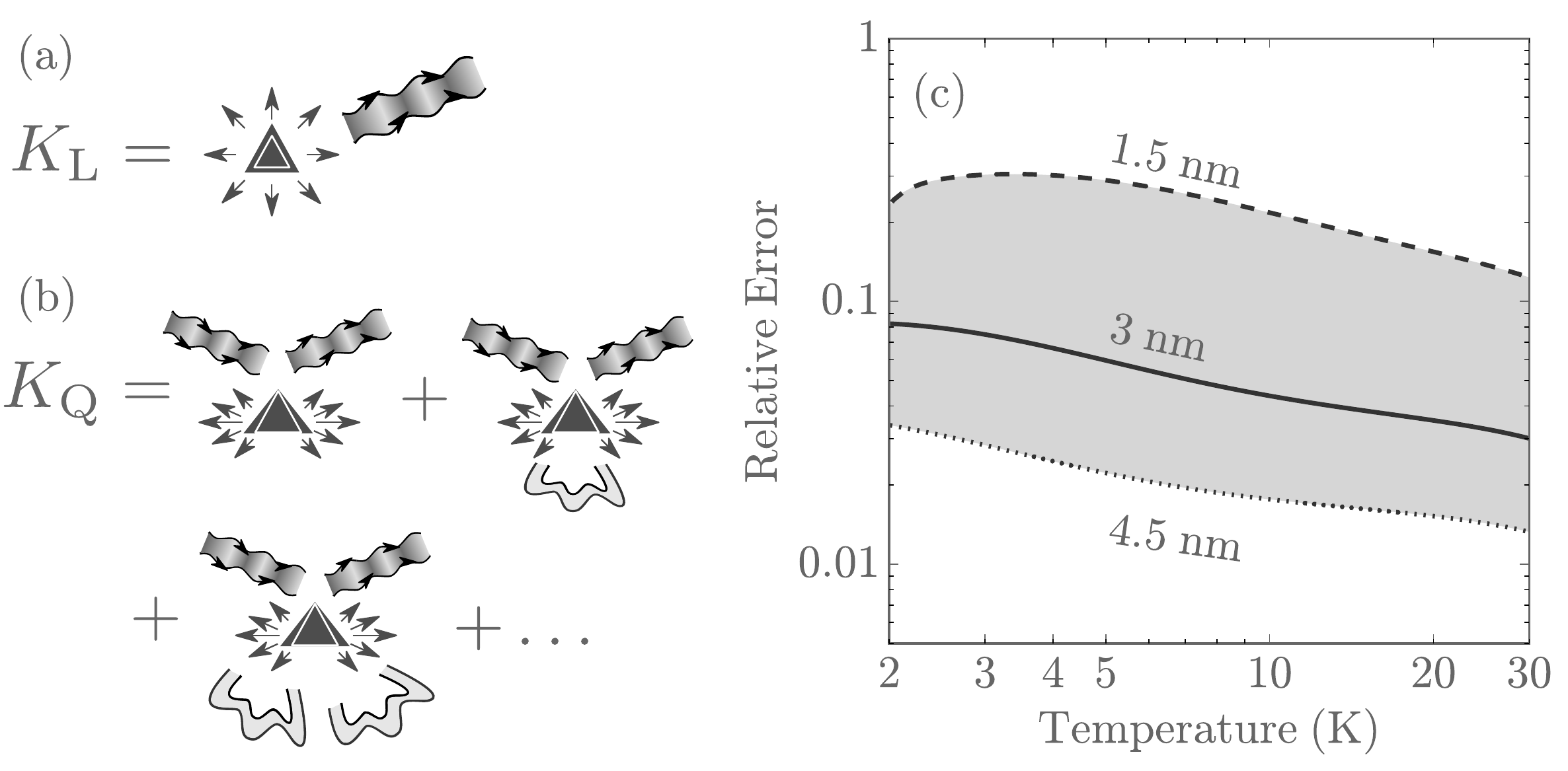}
\caption{ \label{fig:exact_vs_perturbation} (a) The linear cumulant contains a single electron-phonon interaction in the form of an emission or absorption process. (b) The quadratic cumulant contains an infinite series of scattering events. In the present paper we perform the single-scattering approximation meaning that only the first term on the right-hand side is retained. This is equivalent to treating the phonon-induced noise as Gaussian. (c) The relative error in the decay rate, $\eta$, between the exact solution to the infinite series and the single-scattering approximation for $L=\{\SI{1.5}{\nano\meter},\SI{3}{\nano\meter},\SI{4.5}{\nano\meter}\}$.}
\end{figure}

To evaluate the quadratic interaction we consider the triply degenerate excited state only, $l=1$ and $m=\{-1,0,1\}$. Ignoring the higher lying states is justified since $K_\mathrm{Q}\propto\Delta_m^{-2}$. Then the quadratic cumulant, $K_\mathrm{Q}(t)$, becomes diagonal in $m$ in a bulk medium due to the spherical symmetry of the system and leads to
\begin{equation}
K_\mathrm{Q}(t) = \frac{3}{2}\sum_{n=2}^\infty \frac{1}{n}\iiint\ldots\int\mathrm{d}t_1\mathrm{d}t_2\ldots\mathrm{d}t_n D_\mathrm{Q}^{m_am_a}(t_1-t_2)D_\mathrm{Q}^{m_am_a}(t_2-t_3)\ldots D_\mathrm{Q}^{m_am_a}(t_n-t_1),
\end{equation}
where $D_\mathrm{Q}^{m_am_a}(t)$ denotes the quadratic propagator in which $m_a$ can refer to any of the excited states $-1,0,1$. Here we are interested in the long-time decay of $P(t\gg t_0)$ with $t_0=L/v_s$, i.e., the part responsible for the broadening of the zero-phonon line. Evaluating an exact-to-all-order expression for the long-time limit of $K_\mathrm{Q}(t)$ is possible~\cite{skinner86}. Here we discuss the simpler case of the single-scattering approximation in which the QD is assumed to interact with a phonon a single time during the scattering process, see Fig.~\ref{fig:exact_vs_perturbation}(a). In other words, we only consider $n=2$ and neglect the higher order diagrams. 
As explained in the main text, keeping only the $n=2$ term corresponds to treating the phonon bath as Gaussian noise owing to the fundamental property of the cumulant expansion.
This yields
\begin{equation}
K_\mathrm{Q}^\mathrm{3D}(t) \simeq 3\iint\mathrm{d}t_1\mathrm{d}t_2\left[\sum_\qq Q_\qq^{m_a}Q_\qq^{m_a}D_\qq(t_1-t_2) \right]^2,
\end{equation}
which can be evaluated by performing the integration over time
\begin{equation}
\begin{split}
\iint\mathrm{d}t_1\mathrm{d}t_2D_{\qq_1}(t_1-t_2)D_{\qq_2}(t_2-t_1) = -\frac{1}{\hbar^2}&\left\{\left( 2N_{\qq_1}N_{\qq_2} + N_{\qq_1} + N_{\qq_2} + 1 \right)\left[\frac{\sin\left( \frac{\omega_{\qq_1}+\omega_{\qq_2}}{2}t \right)}{\frac{\omega_{\qq_1}+\omega_{\qq_2}}{2}}\right]^2\right.\\
&+ \left.\left( 2N_{\qq_1}N_{\qq_2} + N_{\qq_1} + N_{\qq_2} \right)\left[\frac{\sin\left( \frac{\omega_{\qq_1}-\omega_{\qq_2}}{2}t \right)}{\frac{\omega_{\qq_1}-\omega_{\qq_2}}{2}}\right]^2\right\}.
\end{split}
\label{eq:iint_Dq1_Dq2}
\end{equation}
Using the long-time limit $\omega_\qq^{-2}\sin^2\omega_\qq t \simeq \pi t\delta(\omega_\qq)$ results in a vanishing contribution from the first term on the right-hand side of \eqref{eq:iint_Dq1_Dq2}. We thus obtain $K_\mathrm{Q}^\mathrm{3D}(t) = -\Gamma_\mathrm{3D}t$ with
\begin{equation}
\Gamma_\mathrm{3D} = 3\pi \frac{v_s}{L} C_\mathrm{Q}^2\int_0^\infty \mathrm{d}(qL) (qL)^{10} \textrm{e}^{-(qL)^2} N_q (N_q+1),
\label{eq:gamma_3D_sph}
\end{equation}
where $C_\mathrm{Q} = \left( D_e^2/\Delta_e + D_h^2/\Delta_h \right)/3 (2\pi)^2\rho_m v_s^2 L^3$. In bulk, the quadratic exciton-phonon interaction thus results in a Markovian decay of the coherence. The integrand contains an expression proportional to the squared joint phonon density of states, $q^{10}\exp(-q^2L^2)$, multiplied by a term proportional to the probability to scatter a phonon, $N_q(N_q+1)$. The latter involves a product of the number of incident phonons, $N_q$, and the rate of stimulated emission into a different mode of the same energy, $N_q+1$. 

As opposed to the scattering description, the expression in \eqref{eq:KQ_n_2} can also be seen as the result of the fluctuation in the potential induced by the thermal phonons. These fluctuations are proportional to the variance of the number of phonons in a thermal state, $N_q(N_q+1)$. It is important to emphasize that even though the single-scattering approximation neglects interactions at earlier times, this approximation is very different from the Markov approximation. As we find for instance in 0D structures, the single-scattering approximation allows for highly non-Markovian dynamics. These effects can be interpreted as interferences between single scatterings occurring at different times.

The success of the single-scattering approximation is connected to the weak electron-phonon interaction in QDs, which means that the probability to interact multiple times with the same phonon is negligible. To demonstrate this, we calculate the relative error
\begin{equation}
\eta = \frac{\abs{\Gamma_\mathrm{3D}^\infty-\Gamma_\mathrm{3D}}}{\Gamma_\mathrm{3D}^\infty},
\end{equation}
where $\Gamma_\mathrm{3D}^\infty$ is the dephasing rate that is exact to all orders, and is calculated with the help of Ref.~\cite{skinner86}. Figure~\ref{fig:exact_vs_perturbation}(c) plots $\eta$ as a function of temperature for different QD sizes from which it can be inferred that the single-scattering approximation induces a relative error in the decay rate that is below 10\% for common QD sizes $L\sim\SI{3}{\nano\meter}$.

\section{Maximally confined (0D) nano-structures}
Here we derive the coherence function $P(t)$ for a QD placed at the center of a nano-sphere. The use of open boundary conditions at the surface of the nano-sphere (meaning that the force perpendicular to the surface must vanish so that the sphere vibrates freely) yields two classes of vibrational modes~\cite{masumoto13,auld73}: torsional and spheroidal. The former are purely transversal and do not play any role for deformation-potential interactions. The spheroidal family of modes, $\mathbf{u}_{nlm}$, has three quantum numbers describing the vibrations along the radial direction, $n$, the total angular momentum of the mode, $l$, and its projection along the $z$-axis, $m$. The divergence of the spheroidal mode at the center of the sphere is
\begin{equation}
\divv \mathbf{u}_{nlm}(\rr) = - \mathcal{N}I_{nlm}q_n^2 j_l(q_nr)Y_{lm}(\Omega),
\label{eq:div_u_0D}
\end{equation}
where $j_l$ is the spherical Bessel function of first kind and $l$-th order, $Y_{lm}$ the spherical harmonic, $\Omega$ the solid angle, $I_{nlm}$ the normalization constant of the mode, and $\mathcal{N}=\sqrt{\hbar/2\rho_mv_s}$.
 The linear cumulant is only mediated by modes with $l=m=0$ because the linear exciton-phonon coupling is governed by a matrix element of the form  $\bracket{\psi_{1 b}}{\divv \mathbf{u}_{nlm}}{\psi_{1 b}}$ with $\divv \mathbf{u}_{nlm}$ given in \eqref{eq:div_u_0D}, and the ground-state exciton wavefunction, $\psi_{1b}$, is spherically symmetric. On the other hand, the quadratic cumulant is mediated by modes with $l=1$, $m=\{-1,0,1\}$ because the quadratic exciton-phonon coupling is governed by a matrix element of the form $\bracket{\psi_{1 b}}{\divv \mathbf{u}_{nlm}}{\psi_{2 b}}$, and the excited state, $\psi_{2b}$, has the symmetry of a $p$ orbital. The corresponding families of acoustic modes take the form
\begin{align}
\mathbf{u}_{n00}(\rr) &= -\frac{\mathcal{N}}{2\sqrt{\pi}} I_{n00} j_1(q_n r)\hat{\mathbf{r}},\\
\mathbf{u}_{n1m}(\rr) &= \mathcal{N}\left\{ I_{n1m}\nabla\Psi_{1m}(q_n\mathbf{r}) + J_{n1m}\nabla\times\nabla\times\left[\mathbf{r}\Psi_{1m}(k_n\mathbf{r})\right] \right\},
\end{align}
where $\Psi_{lm}(q\mathbf{r}) = j_l(qr)Y_{lm}(\Omega)$, $k_n = \sigma q_n$, $\sigma = v_s/v_t$ with $v_t$ being the transverse speed of sound, and $\hat{\mathbf{r}}$ is the radial unit vector. $I_{nlm}$ and $J_{nlm}$ contain information about curl-free and divergence-free oscillations of the mode, respectively. Due to the rotational symmetry of the problem, the matrix elements with different angular momenta do not interfere in the quadratic cumulant, see \eqref{eq:KQ_n_2}, resulting in a factorization similar to bulk. This means that it is sufficient to consider a single mode only and just multiply its contribution by a factor of three. In the following we consider the state with $l=1$ and $m=0$. The normalization constants are found to be
\begin{align}
I_{n00}^2 &= \frac{4\tilde{q}_n}{2\left( \cos 2\tilde{q}_n + \tilde{q}_n^2 - 1 \right) + \tilde{q}_n\sin 2\tilde{q}_n},\\
I_{n10}^2 &= \frac{1}{L^2 + r_{n}^2N^2 + 2r_{n}\mathrm{LN}},\\
L^2 &= \frac{1}{4\tilde{q}_n^3}\left[ -4 - 4\tilde{q}_n^2 + 2\tilde{q}_n^4 + 4(1-\tilde{q}_n^2)\cos 2\tilde{q}_n + \tilde{q}_n(\tilde{q}_n^2-8)\sin 2\tilde{q}_n \right],\\
N^2 &= \frac{1}{\sigma^3\tilde{k}_n^3}\left[ -2 - 2\tilde{k}_n^2 + 2\tilde{k}_n^4 + 2(1-\tilde{k}_n^2)\cos 2\tilde{k}_n + \tilde{k}_n(4-\tilde{k}_n^2)\sin 2\tilde{k}_n \right],\\
\mathrm{LN}&= \frac{2}{\tilde{k}_n^3}\left( \tilde{q}_n\cos\tilde{q}_n - \sin\tilde{q}_n \right)\left( \tilde{k}_n\cos\tilde{k}_n - \sin\tilde{k}_n \right),
\end{align}
where $\tilde q = q R$, and $r_{n} = J_{n10}/I_{n10}$. The vibrational eigenfrequencies and $r_n$ are found from applying the open boundary conditions, which yields~\cite{masumoto13}
\begin{align}
-\sigma^2 \tilde{q}_nj_0(\tilde{q}_j) + 4j_1(\tilde{q}_j) &= 0 \textrm{  for  } l=0,\\
\begin{pmatrix}
\alpha_{nlm} & \beta_{nlm} \\
\gamma_{nlm} & \delta_{nlm}
\end{pmatrix}
\begin{pmatrix}
I_{nlm}\\
J_{nlm}
\end{pmatrix}&=0
\textrm{  for  } l>0,\\
\alpha_{nlm} &= -\sigma^2\tilde{q}_n + 2(l+2)j_{l+1}(\tilde{q}_n),\\
\beta_{nlm} &= l\tilde{k}_nj_l(\tilde{k}_n) - 2l(l+2)j_{l+1}(\tilde{k}_n),\\
\gamma_{nlm} &= -\sigma^2\tilde{q}_nj_l(\tilde{q}_n) + 2(l-1)j_{l-1}(\tilde{q}_n),\\
\delta_{nlm} &= (l+1)\left[ 2(l-1)j_{l-1}(\tilde{k}_n) - \tilde{k}_nj_l(\tilde{k}_n) \right].
\end{align}
These equations are solved numerically for each radial mode $n$. 

The relevant electron-phonon matrix elements, $M_{\qq b}^{11}$ and $M_{\qq b}^{12}$, are evaluated for the vibrational mode $(n,l,0)$ as
\begin{align}
\left(M_{\qq b}^{11}\right)_n &= -D_b \frac{\mathcal{N}}{2\sqrt{\pi}}I_{n00}q_n^2\mathrm{e}^{-q_n^2L^2/4},\\
\left(M_{\qq b}^{12}\right)_n &= -D_b \frac{\mathcal{N}}{2\sqrt{\pi}}I_{n10}\frac{q_nL}{\sqrt{6}}q_n^2\mathrm{e}^{-q_n^2L^2/4}.
\end{align}
This allows the computation of the linear and quadratic cumulants. The linear cumulant is evaluated using \eqref{eq:KL_t} as
\begin{equation}
K_\mathrm{L}^{\mathrm{0D}}(t) = -\frac{(D_e-D_h)^2}{8\pi^2\hbar\rho_mv_s^3L^2}\frac{\pi}{2}\sum_n I_{00n}^2(q_nL)^4\mathrm{e}^{-(q_nL)^2/2}\left[v_s^2d_{q_n}(t)\right]
\end{equation}
This expression has a cutoff at wave numbers $q_n\gg L^{-1}$. It is therefore sufficient to evaluate the mechanical frequencies of the nano-sphere up to this cutoff only. We find numerically that the cumulant converges at $q_\mathrm{max}\simeq 5L^{-1}$. The same holds for the quadratic cumulant, which is discussed in the following.

The quadratic propagator is evaluated with the help of \eqref{eq:KQ_general} for the wavefunction with $l=1,m=0$ as
\begin{equation}
D_\mathrm{Q}(t) = \frac{\pi\hbar}{2}C_\mathrm{Q}\sum_n I_{n10}^2q_n^6\mathrm{e}^{-(q_nL)^2/2} D_{q_n}(t).
\end{equation}
As mentioned above, the factorization of the angular momenta and the spherical symmetry result in the same contribution for the states with $l=1,m=\pm 1$. The quadratic cumulant, $K_\mathrm{Q}(t)$, is evaluated numerically in Fig.~3(a) in the main text using the above expression and \eqref{eq:KQ_n_2}, and compared to the $t^2$-approximation. The latter is calculated with the help of \eqref{eq:iint_Dq1_Dq2} by employing the long-time limit $\sin^2[(\omega_{n}-\omega_{n'}) t]/(\omega_n-\omega_{n'})^2\simeq t^2\delta_{nn'}$ yielding
\begin{equation}
\begin{split}
K_\mathrm{Q}^\mathrm{0D}(t) &= - S^2t^2,\\
S^2 &= \frac{3}{2}\left(\frac{\pi}{2}\frac{v_s}{L}C_\mathrm{Q}\right)^2\sum_n I_{n10}^4(q_nL)^{12}\mathrm{e}^{-(q_nL)^2}N_{q_n}\left( N_{q_n}+1 \right).
\end{split}
\end{equation}

\begin{figure}[t!]
\centering
\includegraphics[width=0.8\columnwidth]{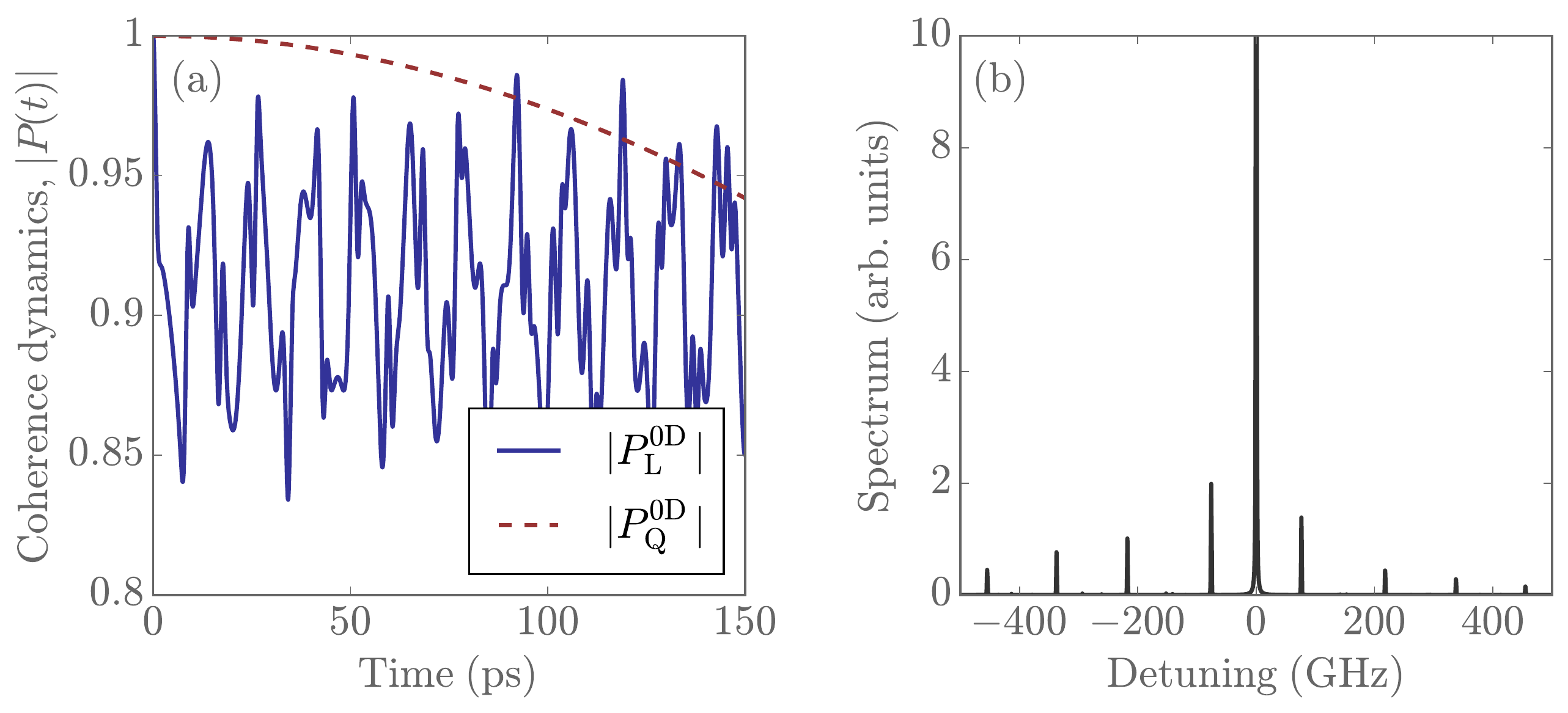}
\caption{ \label{fig:sphere_P_ZPL} (a) Coherence dynamics and (b) emission spectrum of a QD of $L=\SI{3}{\nano\meter}$ embedded in a sphere of $R=\SI{20}{\nano\meter}$ at $T=\SI{5}{\kelvin}$. The spectrum features a ZPL and phonon replicas stemming from the linear exciton-phonon coupling. The lines are broadened by the quadratic electron-phonon coupling. }
\end{figure}

The linear and quadratic coherence functions along with the emission spectrum are plotted in Fig.~\ref{fig:sphere_P_ZPL}. The linear coherence features an interference between the discrete modes of the sphere, which is reflected in the emission spectrum as detuned satellite peaks of the ZPL. Each spectral line is broadened by the quadratic exciton-phonon interaction, and is a Gaussian due to the $t^2$ dependence of the quadratic cumulant, see also Fig.~3 in the main text.

\section{One-dimensional (1D) nano-structures}
\label{sec:1D_structures}
Here we consider the phonon decoherence in realistic 1D photonic waveguides with any cross-sectional shape, which are relevant for various photonic platforms such as nano-wires and nano-beam waveguides.

%Two families of acoustic modes are relevant with a finite $\divv \mathbf{u}$: longitudinal modes in which the oscillations happen solely along the length of the waveguide

The long-time decay of the coherence can be written as a product of a linear term, $P_\mathrm{L0}^\mathrm{1D}$, with low-frequency contributions from the linear exciton-phonon interaction, and a quadratic term, $P_\mathrm{Q}^\mathrm{1D}$, stemming from the quadratic exciton-phonon interaction, see main text for details. As argued in the main text, the quadratic coherence is assumed to be the same as in bulk, $K_\mathrm{Q}^\mathrm{1D}=-\Gamma_\mathrm{3D}t$ due to the large mismatch between the size of the waveguide and of the QD. The evaluation of the linear coherence, $P_\mathrm{L0}^\mathrm{1D}$ is done as follows. Two families of acoustic modes are relevant with a finite $\divv \mathbf{u}$: longitudinal modes in which the oscillations happen solely along the length, $z$, of the waveguide at low energies~\cite{stroscio94}, and flexural modes that bend the waveguide~\cite{auld73}. The former are constant within the cross-section of a 1D waveguide~\cite{LL70} implying that the resulting decoherence is independent of the QD position. This also implies that the dephasing is independent of the shape of the waveguide and only depends on the cross-sectional area, since the phonon energy density is inversely proportional to the latter due to normalization. We thus evaluate the decoherence due to longitudinal modes for a cylindrical waveguide following Ref.~\cite{lindwall07} but note that the result is applicable to any cross-sectional shape with an equivalent area. The vibrational frequencies for a given wave vector $q_z\equiv q$ can be found from applying the traction-free boundary conditions at the surface of the cylinder and solving the corresponding characteristic equations~\cite{stroscio94}. The dispersion of the fundamental mode is found by expanding the characteristic equations in a Taylor series in $q_z\equiv q$ and retaining the lowest orders, which yields $\omega_q = v_\mathrm{1D}q$ with 
\begin{equation}
v_\mathrm{1D} = v_s\sqrt{3+2\nu+\frac{2}{\nu-1}},
\end{equation}
where $\nu$ is the Poisson ratio. Expanding the displacement of the fundamental mode into a Taylor series and retaining the lowest order of $q$ results in
\begin{align}
u_r &= \mathcal{N}_\mathrm{cyl}\nu r\sqrt{q}\mathrm{e}^{\I q z},\\
u_\phi &= 0,\\
u_z &= \frac{\mathrm{i}\mathcal{N}_\mathrm{cyl}}{\sqrt{q}}\mathrm{e}^{\I q z},
\end{align}
where $r$ denotes the radial coordinate, and $\mathcal{N}_\mathrm{cyl}=\sqrt{\hbar/2\rho_mv_\mathrm{1D}AL}$ with $A$ being the cross-sectional area and $L$ the length of the cylinder. The divergence of $\mathbf{u}$ is then to lowest order in $q$
\begin{equation}
\divv \mathbf{u} = -\mathcal{N}_\mathrm{cyl}(1-2\nu)\sqrt{q},
\end{equation}
and is independent of position as expected.
The phonon number is large, $N_q \gg 1$, at such small vibrational frequencies ($\sim\SI{1}{\giga\hertz}$) compared to the thermal energy considered in this work ($\sim\SI{100}{\giga\hertz}$). We can therefore expand $N_q$ in a Taylor series and retain the lowest order in $q$, which is equivalent to the classical equipartition theorem yielding $N_q\hbar\omega_q=k_\mathrm{B}T$. It is worth noting that this expansion may not be valid at much smaller temperatures, a regime that is relevant for other systems such as carbon nano-tubes~\cite{galland08}.
The real part of the linear cumulant can then be evaluated with the help of \eqref{eq:KL_t} to yield to the lowest order in $q$
\begin{equation}
\mathrm{Re}K_\mathrm{L0}^\mathrm{1D}(t) = -\frac{(D_e-D_h)^2(1-2\nu)^2}{4\pi\hbar^2\rho_mv_\mathrm{1D}^2A}k_\mathrm{B}T \int_{-\infty}^\infty \mathrm{d}q \left[\frac{\sin\left(\frac{\omega}{2}t\right)}{\frac{\omega}{2}}\right]^2 \simeq -\frac{(D_e-D_h)^2(1-2\nu)^2k_\mathrm{B}T}{2\hbar^2\rho_mv_\mathrm{1D}^3A}t = - \Gamma_\mathrm{L0}^\mathrm{1D}t,
\label{eq:Gamma_1D_L0}
\end{equation}
where we have again assumed the long-time limit $\sin^2(\omega t)/\omega^2\simeq \pi t \delta(\omega)$. The coherence decay stemming from longitudinal vibrations is therefore fully Markovian in a 1D structure with the rate $\Gamma_\mathrm{1D} = \Gamma_\mathrm{3D} + \Gamma_\mathrm{L0}^\mathrm{1D}$.

\begin{figure}[t!]
\centering
\includegraphics[width=0.4\columnwidth]{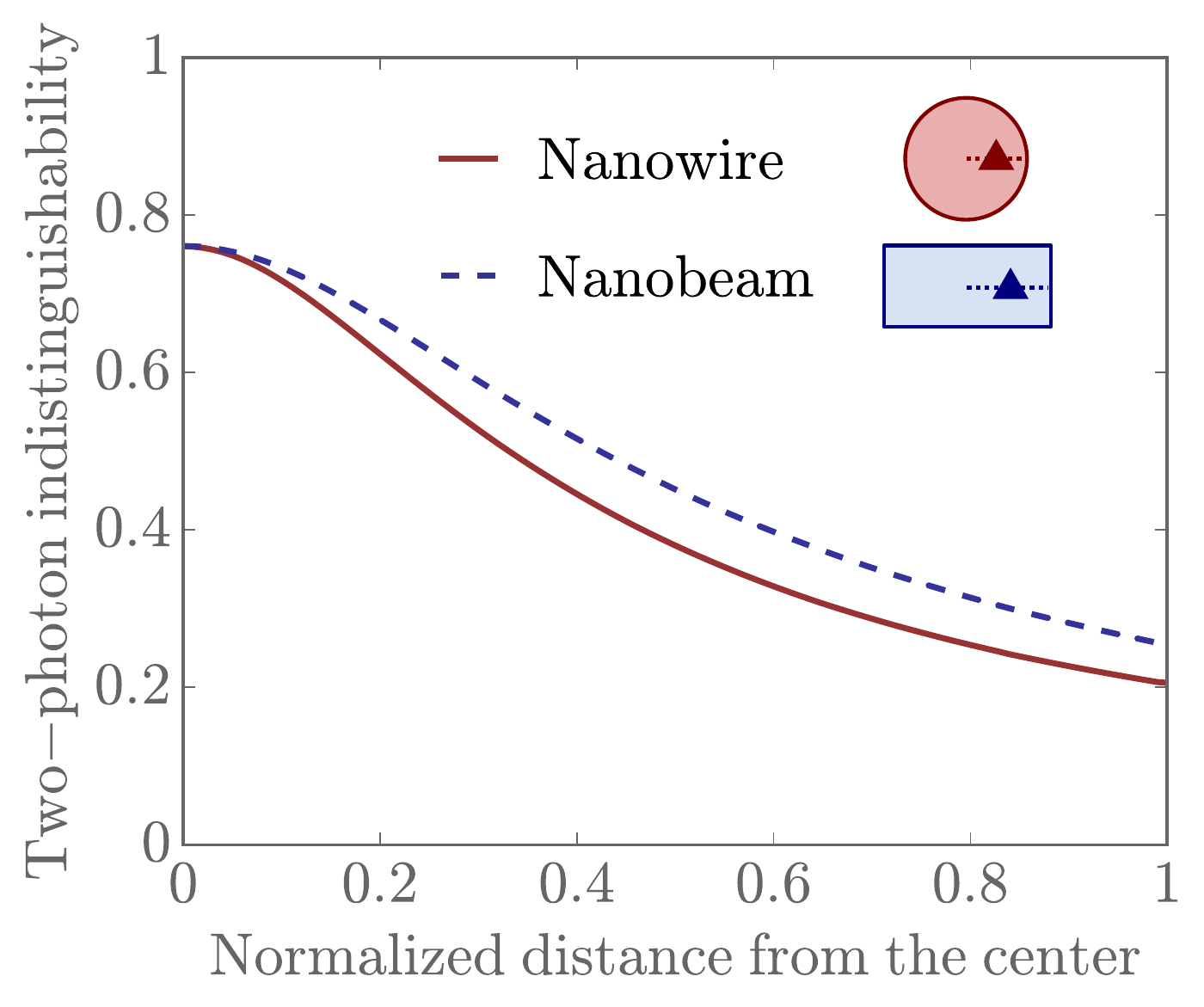}
\caption{ \label{fig:1D_dephasing_vs_position} Indistinguishability between two photons emitted by a QD embedded in a 1D waveguide versus the distance between the QD and the cross-sectional center of the waveguide. The plot shows two waveguides with an equivalent cross-sectional area: a nanowire with a radius of \SI{80}{\nano\meter} (solid red line), and a nanobeam of length \SI{200}{\nano\meter} and width \SI{100}{\nano\meter} (blue dashed line). At the center, only longitudinal modes contribute to the dephasing, and the TPI solely depends on the cross-sectional area and is therefore the same for the two waveguides. Away from the center, the QD also couples to flexural modes leading to a lower photon coherence. A temperature of $T = \SI{5}{\kelvin}$ was considered.}
\end{figure}

Away from the center of the nanowire, the QD is also dephased by flexural modes with a quadratic dispersion at low energies~\cite{LL70} resulting in a non-Markovian decay of coherence, see the main text for details. Figure~\ref{fig:1D_dephasing_vs_position} plots the numerically evaluated dephasing as a function of the QD position within the cross-section of a 1D waveguide. The dephasing due to flexural modes increases significantly as the QD approaches the edge of the waveguide. We have evaluated the decay of coherence due to flexural modes with a similar approach as for longitudinal modes from above, and find that the linear cumulant scales as $\mathrm{Re}K_\mathrm{L0,flexural}^{1D} = -\beta t^{3/2} $ with $\beta > 0$. The total decay of coherence for a QD positioned offcenter is thus $P_\mathrm{1D'} = \exp(-\Gamma_\mathrm{L0}^\mathrm{1D}t-\beta t^{3/2})$, and is plotted in Fig.~3(a) in the main text.

%It turns out that the dephasing rate obtained in \eqref{eq:Gamma_1D_L0} is remarkably robust when changing the shape of the 1D waveguide. In the following we show numerically that the dephasing rate for a rectangular cross-section is the same as for the cylindrical cross-section provided that the areas of the cross-sections are the same.

\section{Two-dimensional (2D) nano-structures} 
\begin{figure}[t!]
\centering
\includegraphics[width=0.8\columnwidth]{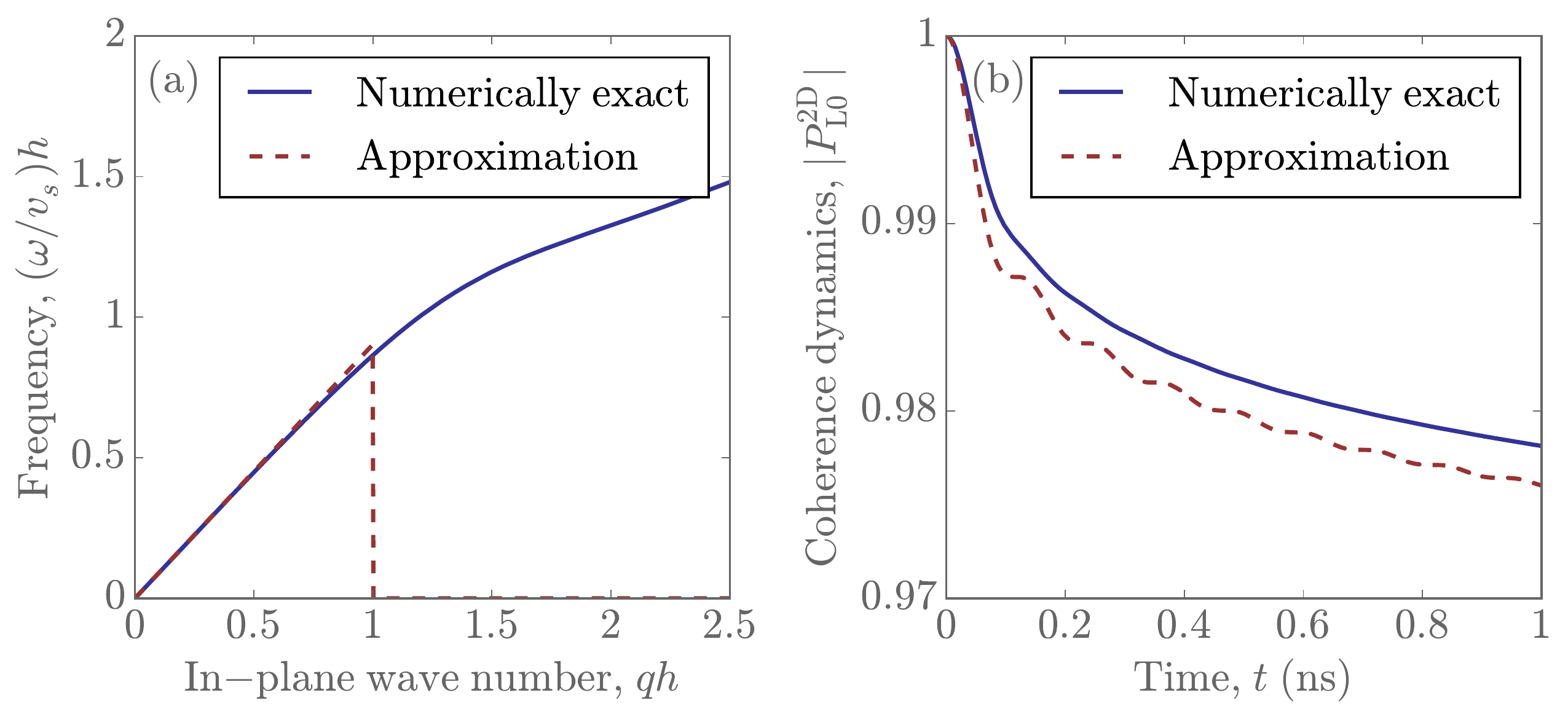}
\caption{ \label{fig:2D_analytics_vs_numerics}  Impact of approximating the dispersion in 2D. Parameters: $h=\SI{80}{\nano\meter}$, $L=\SI{3}{\nano\meter}$, $T=\SI{10}{\kelvin}$. (a) Graphical illustration of the approximation. The numerically exact dispersion is replaced by a linear function with a cutoff at $qh=1$. (b) The approximation allows to obtain an analytic expression for the coherence dynamics, which reproduces with reasonable accuracy the numerically exact solution.}
\end{figure}

Here we evaluate the phonon decoherence of a QD placed in the center of a 2D photonic membrane of realistic thickness. Analogously to the 1D case, the long-time decay of the coherence, $P_\mathrm{ZPL}$, is given by a product of the quadratic coherence function, $P_\mathrm{Q}^\mathrm{2D}$, and the low-frequency contribution to the linear coherence function, $P_\mathrm{L0}^\mathrm{2D}$. The latter is calculated similarly to the case of the cylinder by expanding the fundamental branch of the symmetric Lamb waves (all other solutions are transverse at the QD position) into a Taylor series of the in-plane wave number $q_{||}\equiv q$. The dispersion is linear, $\omega = v_\mathrm{2D}q$, which can be derived by solving the frequency equations~\cite{anghel07} in the low-$q$ and low-$\omega$ limits. The speed of sound in the plane of the membrane, $v_\mathrm{2D}$, is found to be
\begin{equation}
v_\mathrm{2D} = \frac{\sqrt{1-2\nu}}{1-\nu}v_s.
\end{equation}
The general analytic form of the displacement vector, $\mathbf{u}=u_{||}\hat{\mathbf{q}} + u_z \hat{\mathbf{z}}$~\cite{anghel07} ($z$ points along the height of the membrane), is expanded in a Taylor series and normalized to one quantum of energy yielding to lowest order in $q$
\begin{align}
u_{||} &= \I\mathcal{N}_\mathrm{m}\frac{1}{\sqrt{q}}\mathrm{e}^{\I \mathbf{q}\cdot \mathbf{r}_{||}},\\
u_z &= \frac{\nu}{1-\nu}\mathcal{N}_\mathrm{m}z\sqrt{q}\mathrm{e}^{\I \mathbf{q}\cdot \mathbf{r}_{||}},
\end{align}
where $\mathcal{N}_\mathrm{m} = \sqrt{\hbar/2\rho_mv_\mathrm{2D} 2h A}$, $2h$ is the height of the membrane and $A$ the cross-sectional area. The divergence is thus
\begin{equation}
\nabla\cdot\mathbf{u} = -\frac{1-2\nu}{1-\nu}\mathcal{N}_\mathrm{m}\sqrt{q},
\end{equation}
and is again independent of position.
Summing over $q$ yields for the linear cumulant
\begin{equation}
\mathrm{Re}K_\mathrm{L0}^\mathrm{2D}(t) = -\frac{(D_e-D_h)^2(1-2\nu)^2k_\mathrm{B}T}{8\pi \rho_m h v_\mathrm{2D}^4(1-\nu)^2\hbar^2}\int_0^\infty \omega \left[ \frac{\sin\left(\frac{\omega t}{2}\right)}{\frac{\omega}{2}} \right]^2 \mathrm{d}\omega.
\end{equation}
The integral on the right-hand side is diverging, which is caused by the artificial assumption that the dispersion is linear in the entire integration range. In reality, the dispersion becomes sublinear~\cite{anghel07} resulting in a rapidly converging integral for $q \gtrsim h^{-1}$. To obtain an analytic expression, we make the following heuristic assumption for the dispersion, $\omega = v_\mathrm{2D}q,\ \ \ q \leq h^{-1}$, and neglect the modes above $h^{-1}$, see Fig.~\ref{fig:2D_analytics_vs_numerics}(a). A qualitative justification for this assumption is that the acoustic waves become confined to the surface for $q$ larger than $h^{-1}$~\cite{auld73} and do not interact with the QD. We thus obtain for the linear cumulant
\begin{equation}
\mathrm{Re}K_\mathrm{L0}^\mathrm{2D}(t) = -\frac{(D_e-D_h)^2(1-2\nu)^2k_\mathrm{B}T}{4\pi \rho_m h v_\mathrm{2D}^4(1-\nu)^2\hbar^2} \left[ \gamma_\mathrm{E}+\int_{\frac{v_\mathrm{2D}t}{h}}^\infty\mathrm{d}\tau\frac{\cos\tau}{\tau}+\ln\frac{v_\mathrm{2D}t}{h} \right],
\label{eq:K2D_analytic}
\end{equation}
where $\gamma_\mathrm{E}$ is the Euler-Mascheroni constant. At times much larger than $t_\mathrm{m}=h/v_\mathrm{2D}\sim \SI{20}{\pico\second}$ for $h=\SI{80}{\nano\meter}$, this expression is dominated by the last term, $\mathrm{Re}K_\mathrm{L0}^\mathrm{2D} \propto \ln(v_\mathrm{2D}t/h)$, which implies that the linear coherence decays polynomially with time, $P_\mathrm{L0}^\mathrm{2D}(t\gg t_\mathrm{m})=(v_\mathrm{h} t/h)^{-p}$ with $p=0.0085\times T/\SI{1}{\kelvin}$ being the coefficient in front of the right-hand side of \eqref{eq:K2D_analytic}. This analytic result is shown to reproduce qualitatively the numerically exact solution in Fig.~\ref{fig:2D_analytics_vs_numerics}(b). The temporal oscillations occurring with a period of $\sim t_\mathrm{m}$ are artifacts of the truncation: a stepwise truncation in the frequency domain results in a convolution with an oscillatory function in the time domain.

\begin{figure}[t!]
\centering
\includegraphics[width=0.4\columnwidth]{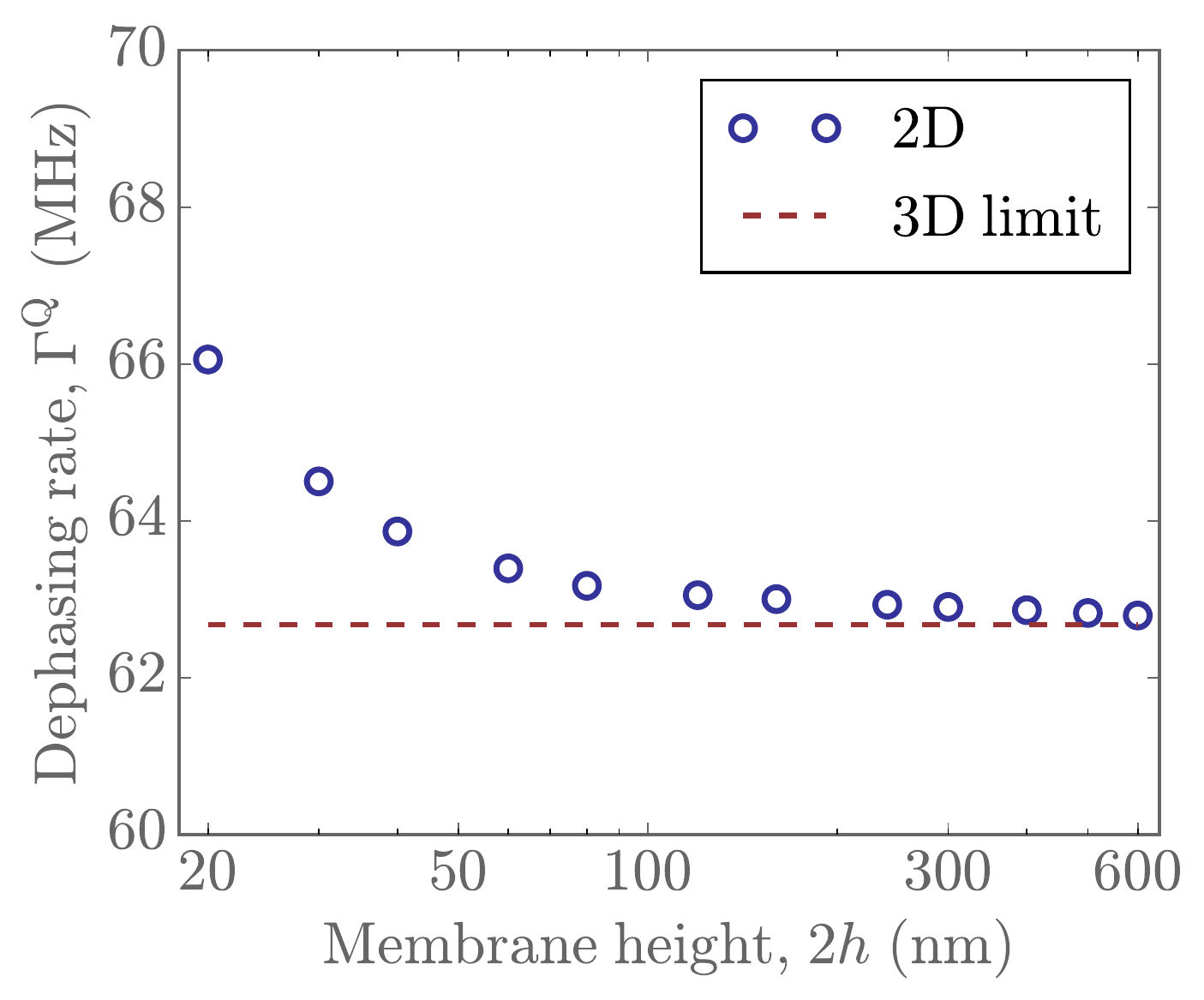}
\caption{ \label{fig:2D_3D_dephasing_rate}  Dephasing due to the quadratic exciton-phonon interaction for a 2D membrane. The rate is found to rapidly converge to the 3D value for realistic membrane thicknesses. Parameters: $L=\SI{3}{\nano\meter}$, $T=\SI{10}{\kelvin}$.}
\end{figure}

The quadratic coherence function, $P_\mathrm{Q}^\mathrm{2D}$, is similar to the bulk value for realistic membrane thicknesses as argued in the main text. To demonstrate this quantitatively, we evaluate the quadratic interaction in \eqref{eq:KQ_n_2} for a QD placed in the center of the membrane. All the modes up to $q=5/L$ and $\omega/v_s=5/L$ are evaluated numerically~\cite{anghel07}. It is found that the propagators with different orbital symmetry do not interfere, $D_\mathrm{Q}^{mn} = D_\mathrm{Q}^{mn}\delta_{mn}$. This is due to the rotational symmetry around the $z$-axis meaning that the phonon modes can still be characterized by the angular momentum quantum number pointing along $z$. However, compared to bulk, the three terms are no longer equal: $D_\mathrm{Q}^{m=-1,n=-1}=D_\mathrm{Q}^{m=+1,n=+1}\neq D_\mathrm{Q}^{m=0,n=0}$, where in this case $m$ and $n$ denote the orbital angular momentum of the QD state. The terms $D_\mathrm{Q}^{m=n=\pm 1}$ are described by the interaction with symmetric (with respect to the center of the membrane) Lamb waves, $\mathbf{u}_s$, while $D_\mathrm{Q}^{m=n=0}$ is mediated by anti-symmetric vibrations, $\mathbf{u}_a$~\cite{anghel07}. The divergence of the modes is found to be
\begin{align}
\divv \mathbf{u}_s &= - \mathcal{N}_\mathrm{m}N_s (q^2+\alpha^2)A \cos(\alpha z)\mathrm{e}^{\I \mathbf{q}\cdot \mathbf{r}_{||}},\\
\divv \mathbf{u}_a &= \mathcal{N}_\mathrm{m}N_a (q^2+\alpha^2)B \sin(\alpha z)\mathrm{e}^{\I \mathbf{q}\cdot \mathbf{r}_{||}},
\end{align}
where $\alpha$ and $\beta$ are the transverse wavevectors subject to $\alpha^2=\sqrt{\omega^2/v_s^2 - q^2}$ and $\beta^2=\sqrt{\omega^2/v_t^2 - q^2}$, respectively, $A=2\beta q\cos\beta$, $B = -2\beta q \sin\beta$, $v_t$ is the transverse speed of sound, and $N_s$ and $N_a$ are normalization factors given in Ref.~\cite{anghel07}. The electron-phonon matrix elements are evaluated as
\begin{align}
M_{\mathbf{q}b}^{12_{m=\pm 1}} &= -D_b\mathcal{N}_\mathrm{m}\frac{\I}{\sqrt{2}}N_sAL(\pm q_x+\I q_y)(q^2+\alpha^2)\mathrm{e}^{-(q^2+\alpha^2)L^2/4},\\
M_{\mathbf{q}b}^{12_{m=0}} &= D_b\mathcal{N}_\mathrm{m}\frac{1}{\sqrt{2}}N_aBL\alpha (q^2+\alpha^2)\mathrm{e}^{-(q^2+\alpha^2)L^2/4}.
\end{align}
The propagators are therefore found to be
\begin{align}
D_\mathrm{Q}^{m=n=\pm 1} &= \frac{3\pi h}{L}C_Q \sum_{\alpha}\int_0^\infty q^3 (\alpha^2+q^2)^{3/2} \abs{N_s}^2 \mathrm{e}^{-(q^2+\alpha^2)/2} \hbar D_\mathbf{q}(t)\mathrm{d}q,\\
D_\mathrm{Q}^{m=n=0} &= \frac{6\pi h}{L}C_Q \sum_{\alpha}\int_0^\infty q \alpha^2 (\alpha^2+q^2)^{3/2} \abs{N_a}^2 \mathrm{e}^{-(q^2+\alpha^2)/2} \hbar D_\mathbf{q}(t)\mathrm{d}q.
\end{align}
The resulting decay rate is found to be Markovian with the rate $\Gamma_\mathrm{2D}^\mathrm{Q}$, which is compared to the bulk value, $\Gamma_\mathrm{3D}$, in Fig.~\ref{fig:2D_3D_dephasing_rate} as a function of membrane thickness. The 2D rate is very close to the bulk limit for realistic membrane thicknesses, which is why the quadratic interaction is assumed to be bulk-like throughout the main text.

\end{document}